\documentclass[11pt]{article}

\usepackage{amsmath}
\usepackage{graphicx}
\usepackage{amsfonts}
\usepackage{amssymb}
\usepackage{epsfig}
\usepackage{color}
\usepackage{psfrag}
\usepackage{epstopdf}
\usepackage{mathrsfs}

\newcommand{\EQ}{\begin{equation}}
\newcommand{\EN}{\end{equation}}
\newcommand{\be}{\begin{equation}}
\newcommand{\ee}{\end{equation}}
\newcommand{\bea}{\begin{eqnarray}}
\newcommand{\eea}{\end{eqnarray}}

\newcommand{\rd}{{\rm d}}

\usepackage{bm}
\newcommand{\fig}[1]{\parbox{2.0cm}{\epsfig{file=#1,height=1.8cm}}}

\newcommand{\newfig}[1]{\parbox{1.8cm}{\epsfig{file=#1,height=3.4cm}}}

\begin{document} \setcounter{page}{0}
\topmargin 0pt
\oddsidemargin 5mm
\renewcommand{\thefootnote}{\arabic{footnote}}
\newpage
\setcounter{page}{0}
\topmargin 0pt
\oddsidemargin 5mm
\renewcommand{\thefootnote}{\arabic{footnote}}

\newpage

\begin{titlepage}
\begin{flushright}
\end{flushright}
\vspace{0.5cm}
\begin{center}
{\large {\bf Exact theory of intermediate phases in two dimensions}}\\
\vspace{1.8cm}
{\large Gesualdo Delfino and Alessio Squarcini}\\
\vspace{0.5cm}
{\em SISSA -- Via Bonomea 265, 34136 Trieste, Italy}\\
{\em INFN sezione di Trieste}\\
\end{center}
\vspace{1.2cm}

\renewcommand{\thefootnote}{\arabic{footnote}}
\setcounter{footnote}{0}

\begin{abstract}
\noindent
We show how field theory yields the exact description of intermediate phases in the scaling limit of two-dimensional statistical systems at a first order phase transition point. The ability of a third phase to form an intermediate wetting layer or only isolated bubbles is explicitly related to the spectrum of excitations of the field theory. The order parameter profiles are determined and interface properties such as passage probabilities and internal structure are deduced from them. The theory is illustrated through the application to the $q$-state Potts model and Ashkin-Teller model. The latter is shown to provide the first exact solution of a bulk wetting transition.
\end{abstract}
\end{titlepage}

\newpage
\section{Introduction}
\begin{figure}[htbp]
\centering
\includegraphics[width=5cm]{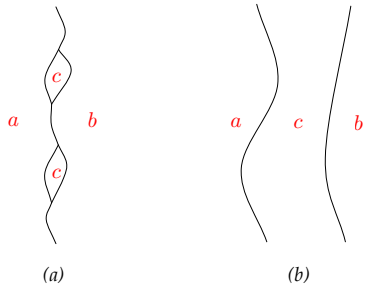}
\caption{Two different regimes of phase separation: a third phase appears in bubbles (a), or through a wetting layer (b).}
\label{wetting}
\end{figure}

Statistical systems at a first order phase transition point allow for phase coexistence. Boundary conditions can be chosen to select a phase $a$ on the left half of the system and a phase $b$ on the right half, the two phases being separated by an interfacial region whose characterization is a particularly interesting problem. In systems allowing for a third degenerate phase, the latter can appear in the interfacial region either via the formation of small bubbles (or drops, Fig.~\ref{wetting}a), or because a macroscopic (``wetting'') intermediate layer of phase $c$ forms between phases $a$ and $b$ (Fig.~\ref{wetting}b). A transition from the first to the second regime induced by the variation of a parameter of the system goes under the name of wetting transition (see e.g. \cite{Dietrich}).

The physics of phase separation is known to be sensitive to dimensionality. The two-dimensional case, in particular, possesses specific features originating from especially strong fluctuations of the interfaces. A key role in establishing the existence of these peculiarities was played by exact results for the planar Ising model \cite{Abraham}, which then were used to test the reliability of heuristic descriptions (see in particular \cite{Fisher}). While the technical complexity of lattice derivations has restricted them to the Ising case, field theory should provide the natural framework for a general study of universal properties in the scaling limit. Nonetheless, a field theory of phase separation in two dimensions has been missing, arguably because the aforementioned peculiarities involve field theoretical counterparts. Only recently it has been shown \cite{DV,DS} how the Ising results follow as particular cases of the general and exact field theoretical formalism which consistently takes into account the fact that interfaces in two dimensions correspond to trajectories of topological excitations (kinks) propagating in imaginary time.

In this paper we extend the formalism of \cite{DV} to study systems with a third phase in both regimes of Fig.~\ref{wetting} and the wetting transition. These systems are not of Ising type and have not been studied previously in a direct and exact way. Clearly, a main point is the characterization of the notion of interface. Being extended, interfaces are not fundamental objects of field theory. Hence we have to deduce their statistical properties from the determination of the spatial dependence of the order parameter, which is the expectation value of a local operator and the indicator of phase separation. We show how this analysis can be carried out in general in field theory and how it is intimately related to the connectedness properties of matrix elements on kink states. 

We derive in particular the following properties. Whether the third phase is wetting or not is determined by the spectrum of kinks of the field theory. The interfacial tension between two phases coincides with the mass of the lightest kink connecting these two phases, and the equilibrium condition among the three interfacial tensions at the vertex of a bubble coincides with energy conservation for the relativistic particles at a bound state vertex. The transverse fluctuations of the interface in the non-wetting regime of Fig.~\ref{wetting}a are Gaussian with a width increasing as $R^{1/2}$, where $R$ is the size of the system in the direction parallel to the interface; the size in the transverse direction is assumed infinite, while $R$ is taken much larger than the correlation length in the pure phases, which in turn is inversely proportional to the mass scale. The effect on the order parameter of the bubbles of Fig.~\ref{wetting}a vanishes as $R^{-1/2}$ to leave a sharp separation between phases $a$ and $b$ in the asymptotic large $R$ limit. For systems in which the external phases are exchanged by a symmetry, the coefficient of this bubble term depends only on the bulk theory and can also be determined exactly in many cases. The subsequent term in the large $R$ expansion corresponds to trifurcations rather than bifurcations in Fig.~\ref{wetting}a and is suppressed as $R^{-1}$; in two-phase, Ising-like systems this provides the first correction to sharp separation. In the wetting regime of Fig.~\ref{wetting}b the order parameter profile does not approach at large $R$ that corresponding to sharp separation between phases $a$ and $b$. Its exact determination leads to a combined passage probability which differs from that of two independent interfaces by a factor of the square of the distance between the interfaces, which are then mutually avoiding. The transition between the two regimes corresponds to the unbinding of a bound state and we exhibit the Ashkin-Teller model as a first exactly solved example of such a bulk wetting transition.

The paper is organized as follows. In the next section we review the results of \cite{DV} for the non-wetting regime. The field theoretical formalism for the wetting case is then developed in section~3, and specialized to the $q$-state Potts model and to the Ashkin-Teller model in sections~4 and 5, respectively. Few conclusive remarks are given in section~6, while an appendix contains the evaluation of some integrals.

\section{Adjacent phases and single interfaces}
In this section we review the field theoretical framework of \cite{DV} for the study of phase separation in two-dimensions and the results for the case in which there is not formation of an intermediate wetting phase. To be definite we refer to a spin model with short range ferromagnetic interactions, at a first order phase transition point. The spin variable can take discrete values labelled by an integer $a=1,\dots,n$, and the system can be brought into a pure (translationally invariant) phase of type $a$ fixing the boundary spins to the value $a$ and then removing the boundary to infinity.

We consider the scaling limit close to a continuous phase transition point, which yields a Euclidean field theory on the plane with coordinates $(x,y)$. Such a theory amounts to the continuation to imaginary time $t = iy$ of a relativistic field theory in one space dimension. Phase coexistence corresponds in the relativistic theory to the presence of degenerate vacua $|\Omega_{a}\rangle$ associated to the pure phases of the system. In $1+1$ dimensions the elementary excitations are stable kink states $|K_{ab}(\theta)\rangle$ which interpolate between two different vacua $|\Omega_{a}\rangle$ and $|\Omega_{b}\rangle$. These topological excitations are relativistic particles with enegy-momentum 
\EQ
(e,p)=m_{ab} \left( \cosh\theta,\sinh\theta \right)\,,
\label{rapidity}
\EN
where $\theta$ is called rapidity and $m_{ab}$ is the kink mass. Two vacua (phases) are not necessarily connected by an elementary kink, and in this case we call them non-adjacent; non-adjacent vacua will be connected by a multi-kink excitation $\vert K_{av_{1}}(\theta_{1})K_{v_{1}v_{2}}(\theta_{2}) \dots K_{v_{n-1}b}(\theta_{n})\rangle$ which visits other vacua along the way.

\begin{figure}[htbp]
\centering
\includegraphics[width=12cm]{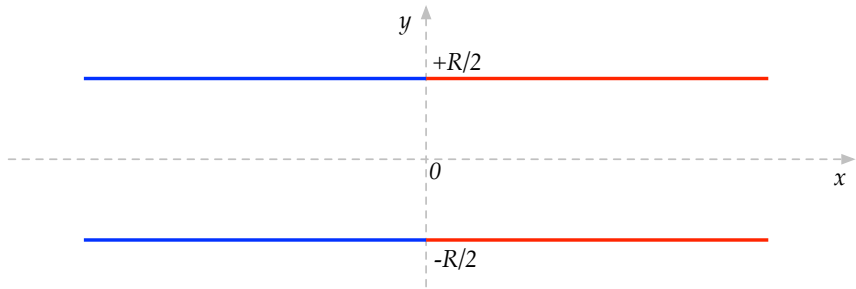}
\caption{$ab$ boundary conditions: the boundary spins are fixed to the value $a$ for $x<0$ and to a different value $b$ for $x>0$. We will denote by $\langle\sigma(x,y)\rangle_{ab}$ the magnetization on the strip with these boundary conditions.}
\label{rectangle}
\end{figure}

We now consider the system on a horizontal strip of width $R$ and fix the boundary spins to a value $a$ for $x<0$ and to a value $b\neq a$ for $x>0$ ($ab$ boundary conditions, Fig.~\ref{rectangle}). Phase separation is expected to emerge when $R$ becomes much larger than the correlation length of the pure phases, which is inversely proportional to $m_{ab}$. In this section we review the case of separation between adjacent phases $a$ and $b$ studied in \cite{DV}.

The boundary condition at time $t$ switching from $a$ to $b$ at $x=x_{0}$ is implemented by a boundary state $\vert B_{ab}(x_{0};t)\rangle$ which can be decomposed over the complete basis of states of the bulk theory (the kink states). Since the states entering the decomposition have to interpolate between the phases $a$ and $b$ and the latter are adjacent, we have 
\EQ
\label{single01}
\vert B_{ab}(x_{0};t) \rangle = \textrm{e}^{-itH+ix_{0}P} \biggl[ \int_{\mathbb{R}} \frac{\rd \theta}{2\pi} \, f_{ab}(\theta) \, \vert K_{ab}(\theta) \rangle +\dots \biggr],
\EN
where $H$ and $P$ are the energy and momentum operators of the $(1+1)$-dimensional theory, and the dots correspond to states with total mass larger than $m_{ab}$. The partition function on the strip with $ab$ boundary conditions then reads\footnote{Kink states are normalized by $\langle K_{ab}(\theta)|K_{b'a'}(\theta')\rangle=2\pi\delta_{aa'}\delta_{bb'}\delta(\theta-\theta')$. In (\ref{single02}) and below the symbol $\simeq$ referred to functions of $R$ indicates omission of terms subleading for $R$ large. }
\bea
\mathcal{Z}_{ab}(R) &=& \langle B(x_{0};iR/2) \vert B(x_{0};-iR/2) \rangle\nonumber\\
&\simeq& \int_{\mathbb{R}} \rd \theta \, |f_{ab}(\theta)|^{2} \, \textrm{e}^{-m_{ab}R \cosh\theta} \simeq \frac{|f_{ab}(0)|^{2} \, \textrm{e}^{-m_{ab}R}}{\sqrt{2\pi m_{ab}R}},
\label{single02}
\eea
where in the second line we took the large $R$ limit which projects onto the lightest (single kink) state\footnote{The minimal energy of an asymptotic state is its total mass.} in (\ref{single01}) and makes the integral dominated by small rapidities. Phase separation amounts to the creation of two pure phases on the far left and on the far right, separated by an interfacial region. The excess free energy due to the creation of the interface divided by $R$ is called interfacial tension and corresponds to
\EQ
\label{single04}
\Sigma_{ab} = -\lim_{R\rightarrow\infty}\frac{1}{R}\ln\frac{\mathcal{Z}_{ab}(R)}{\mathcal{Z}_{a}(R)},
\EN
where $\mathcal{Z}_{a}(R)$ is the partition function with all the boundary spin fixed to the value $a$. The corresponding boundary state expands over bulk states interpolating between $a$ and $a$, the lightest of which is the vacuum $|\Omega_a\rangle$, so that for large $R$ we have $\mathcal{Z}_{a}(R)\simeq\langle \Omega_{a} \vert \Omega_{a} \rangle=1$; (\ref{single02}) then yields 
\EQ
\Sigma_{ab}=m_{ab}\,.
\label{tension}
\EN

The local magnetization at a point $(x,y)$ on the strip with $ab$ boundary conditions reads
\EQ
\label{single05}
\langle \sigma(x,y) \rangle_{ab} = \frac{ \langle B_{ab}(0;iR/2) \vert \sigma(x,y) \vert B_{ab}(0;-iR/2) \rangle }{ \langle B_{ab}(0;iR/2) \vert B_{ab}(0;-iR/2) \rangle },
\EN
where $\sigma(x,y)$ is the magnetization operator, satisfying 
\EQ
\label{single07}
\sigma(x,y) = \textrm{e}^{ixP+yH} \sigma(0,0) \textrm{e}^{-ixP-yH}.
\EN
Use of the boundary state (\ref{single01}) gives
\EQ
\label{single06}
\langle \sigma(x,y) \rangle_{ab} = \frac{1}{\mathcal{Z}_{ab}(R)} \int_{\mathbb{R}^{2}} \frac{\rd \theta}{2\pi} \frac{\rd \theta^{\prime}}{2\pi} \, f_{ab}^{*}(\theta) f_{ab}(\theta^{\prime}) \, \langle K_{ab}(\theta) \vert \, \textrm{e}^{(-\frac{R}{2}+y)H} \sigma(x,0) \textrm{e}^{-(\frac{R}{2}+y)H} \, \vert K_{ab}(\theta^{\prime}) \rangle + \dots;
\EN
the dots stay for the contribution coming from states with higher mass, which for any fixed $|y|<R/2$ becomes negligible as $R\to\infty$. Then in this limit we have
\EQ
\label{single08}
\langle \sigma(x,y) \rangle_{ab} \simeq \frac{1}{\mathcal{Z}_{ab}(R)} \int_{\mathbb{R}^{2}} \frac{\rd \theta}{2\pi} \frac{\rd \theta^{\prime}}{2\pi} \, f_{ab}^{*}(\theta) f_{ab}(\theta^{\prime}) \, \mathcal{M}_{ab}^{\sigma}(\theta \vert \theta^{\prime}) \, \mathcal{O}(\theta,\theta^{\prime}),
\EN
where\footnote{From now on we will most of the times drop the indices on the kink mass to simplify the notation.} $\mathcal{O}(\theta,\theta^{\prime})  = \textrm{e}^{-M_{-}\cosh\theta-M_{+}\cosh\theta^{\prime}} \, \textrm{e}^{imx \left( \sinh\theta-\sinh\theta^{\prime}\right)}$, $M_{\pm} = m(R/2 \pm y)$, and
\EQ
\label{single09}
\mathcal{M}_{ab}^{\sigma}(\theta \vert \theta^{\prime}) \equiv \langle K_{ab}(\theta) \vert \, \sigma(0,0) \, \vert K_{ba}(\theta^{\prime}) \rangle.
\EN
The matrix element (\ref{single09}) decomposes as 
\EQ
\label{single10}
\mathcal{M}_{ab}^{\sigma}(\theta \vert \theta^{\prime}) = \mathscr{F}_{ab}^{\sigma,R}(\theta \vert \theta^{\prime})+ 2\pi \delta(\theta-\theta^{\prime}) \langle \sigma \rangle_{a},
\EN
into the sum of a connected and a disconnected part; $\langle \sigma \rangle_{a}$ denotes the magnetization in the pure phase $a$. Such a decomposition corresponds to the pictorial representation
\EQ
\newfig{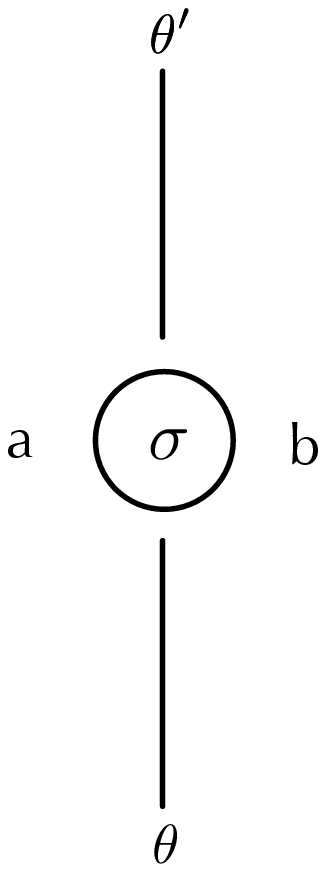} = \newfig{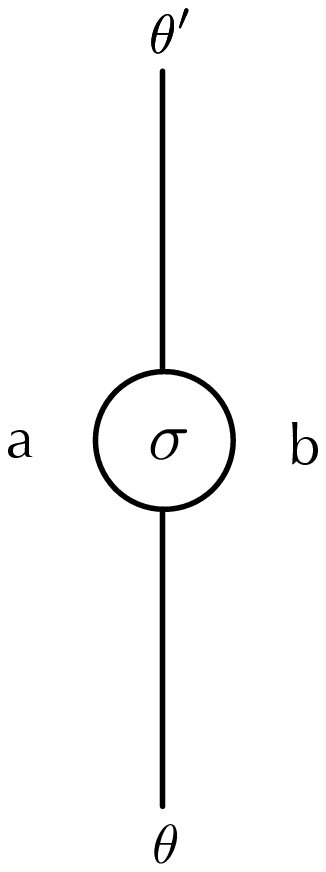} + \newfig{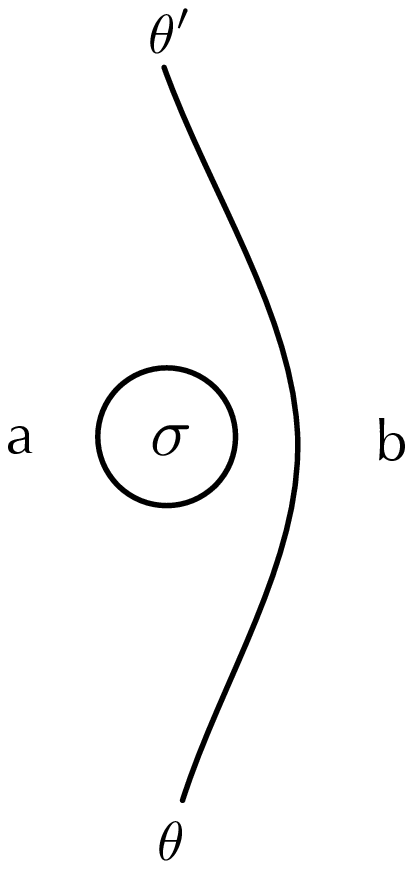}, \hspace{1cm}
\EN
where the disconnected trajectory passes to the right of the insertion point of the magnetization operator, which is then evaluated in the phase $a$. Of course, the decomposition in which the disconnected trajectory passes to the left of the insertion point is also allowed, and in this case $\mathscr{F}_{ab}^{\sigma,L}(\theta \vert \theta^{\prime})$ and $\langle \sigma \rangle_{b}$ replace $\mathscr{F}_{ab}^{\sigma,R}(\theta \vert \theta^{\prime})$ and $\langle \sigma \rangle_{a}$ in (\ref{single10}). It follows that $\mathscr{F}_{ab}^{\sigma,R}(\theta \vert \theta^{\prime})$ and $\mathscr{F}_{ab}^{\sigma,L}(\theta \vert \theta^{\prime})$ coincide for $\theta\neq\theta'$, while for $\theta\to\theta'$ they behave as\footnote{Relativistically invariant quantities depend on rapitity differences.}
\EQ
i\frac{\langle \sigma \rangle_{a}-\langle \sigma \rangle_b}{\theta-\theta'\mp i\epsilon}\,,
\label{pole}
\EN
with the upper (resp. lower) sign referring to $\mathscr{F}_{ab}^{\sigma,R}(\theta \vert \theta^{\prime})$ (resp. $\mathscr{F}_{ab}^{\sigma,L}(\theta \vert \theta^{\prime})$). For the purpose of generalization in subsequent sections we use the pictorial representation\footnote{Kinematical poles like (\ref{pole}) are well known to experts of two-dimensional integrable field theory (see \cite{BKW,Smirnov} and, for the case of kink excitations of interest here, \cite{DC98,DV10}). While integrability simplify the scattering theory and allows the general determination of residues, kinematical poles exist in any two-dimensional field theory. For the two-leg case (\ref{residue}) no scattering is involved and the residue is completely general.}
\EQ
-i \, \textrm{Res}_{\theta_{1}=\theta_{2}} \newfig{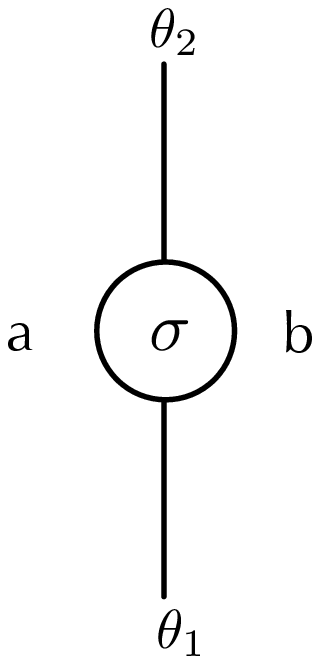} = \newfig{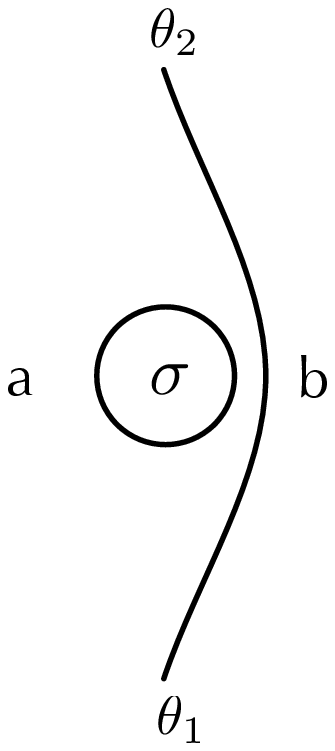} - \quad\newfig{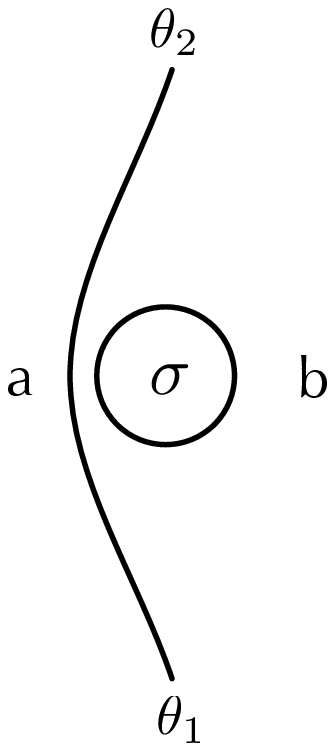} = \langle \sigma \rangle_{a} - \langle \sigma \rangle_{b}\,.
\label{residue}
\EN

\begin{figure}[htbp]
\centering
\includegraphics[width=8cm]{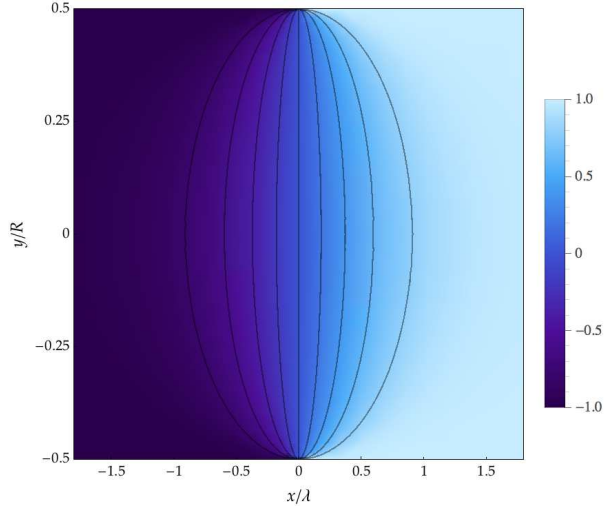}
\caption{Ising magnetization $\langle \sigma(x,y) \rangle_{-+}/\langle\sigma\rangle_{+}$. The ellipses correspond to constant values of $\chi$, and then to constant values of the magnetization.}
\label{ising_1}
\end{figure}

Once we substitute (\ref{single10}) into (\ref{single08}) and take into account that for large $R$ the integral is dominated by $\theta\simeq\theta'\simeq 0$, we can use (\ref{pole}) to obtain
\EQ
\langle \sigma(x,y) \rangle_{ab}\simeq\langle \sigma \rangle_{a}+ \frac{\vert f_{ab}(0) \vert^{2} \, \textrm{e}^{-mR}}{\mathcal{Z}_{ab}(R)} \int_{\mathbb{R}^{2}} \frac{\rd \theta}{2\pi} \frac{\rd \theta^{\prime}}{2\pi} \, \frac{i\Delta \langle \sigma\rangle}{\theta-\theta^{\prime}-i\epsilon} \, \textrm{e}^{-\bigl[ \frac{M_{-}}{2}\theta^{2} + \frac{M_{+}}{2} ( \theta^{\prime} )^{2} \bigr] + imx (\theta-\theta^{\prime})},
\EN
where $\Delta \langle \sigma \rangle\equiv\langle \sigma \rangle_{a} - \langle \sigma \rangle_{b}$. Defining $\theta_{\pm} = \sqrt{mR/8} \, ( \theta \pm \theta^{\prime} )$, 
\EQ
\lambda\equiv\sqrt{R/(2m)}\,,
\label{lambda}
\EN
\EQ
\kappa\equiv\sqrt{1-4y^2/R^2}\,,\hspace{1cm}|y|<\frac{R}{2}\,,
\label{kappa}
\EN
and performing the integral on $\theta_+$ gives
\EQ
\label{single14}
\langle \sigma(x,y) \rangle_{ab}=\langle \sigma \rangle_{a} + \frac{i\Delta \langle \sigma \rangle}{2\pi} \int_{\mathbb{R}} \frac{\rd \theta_{-}}{\theta_{-}-i\epsilon} \, \textrm{e}^{-\kappa^2\theta_{-}^{2}+2ix\theta_{-}/\lambda}\,.
\EN
We can now differentiate with respect to $x$ in order to cancel the pole, perform the Gaussian integral over $\theta_-$ and integrate back over $x$ with the asymptotic condition $\langle\sigma(+\infty,y)\rangle_{ab}=\langle\sigma\rangle_{b}$; the result is 
\EQ
\label{single17}
\langle \sigma(x,y) \rangle_{ab}\simeq\frac{\langle \sigma \rangle_{a} + \langle \sigma \rangle_{b}}{2} - \frac{\langle \sigma \rangle_{a} - \langle \sigma \rangle_{b}}{2} \, \textrm{erf}(\chi)\,,
\EN
where $\textrm{erf}(x)=(2/\sqrt{\pi})\int_{0}^{x} \rd u \, \textrm{e}^{-u^{2}}$ is the error function and  
\EQ
\chi\equiv \frac{x}{\lambda\kappa}\,.
\label{chi}
\EN
For the Ising model $\langle \sigma \rangle_{a}=-\langle \sigma \rangle_{b}=\langle \sigma \rangle_{\pm}$ and (\ref{single17}) reduces to $-\langle \sigma \rangle_{\pm}\textrm{erf}(\chi)$, which is the scaling limit of the exact lattice result of \cite{Abraham81}. The Ising magnetization $\langle \sigma(x,y) \rangle_{-+}$ is shown in Fig.\ref{ising_1}.

Subleading terms in the large $R$ expansion of (\ref{single17}) can be worked out systematically from the smal rapidity expansion of the boundary amplitude $f_{ab}(\theta)$ and of the matrix element $\mathscr{F}_{ab}^\sigma(\theta|\theta')$. If the phases $a$ and $b$ play a symmetric role, we have $f_{ab}(\theta)=f_{ab}(0)+O(\theta^2)$, and the next contribution to (\ref{single17}) is easily found to be
\EQ
\frac{\mathscr{C}_{ab}^\sigma}{m}\,\frac{\textrm{e}^{-\chi^2}}{\sqrt{\pi}\,\kappa\lambda}\,,
\label{branching}
\EN
where $\mathscr{C}_{ab}^\sigma$ is the coefficient $c_0$ of the expansion $\mathscr{F}_{ab}^\sigma(\theta|\theta')=\sum_{k=-1}^\infty c_k(\theta-\theta')^k$, and then depends only on the bulk theory.

\begin{figure}[htbp]
\centering
\includegraphics[width=9cm]{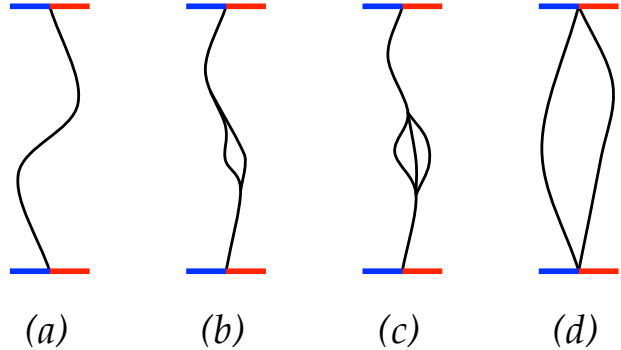}
\caption{Some configurations of a single interface (a,b,c) and the leading large $R$ configuration of a double interface (d).}
\label{configurations}
\end{figure}
It is easy to see that the result (\ref{single17}) corresponds to the average over the configurations of an interface which intersects only once the lines $y=\textrm{constant}$ and sharply separates two pure phases $a$ and $b$ (Fig.~\ref{configurations}a). Indeed, if we call $p(u,y)du$ the probability that such an interface passes in the interval $(u,u+du)$ on the line of constant $y$ on the strip, and\footnote{We denote by $\theta(x)$ the step function which equals 1 if $x>0$ and $0$ if $x<0$.} 
\EQ
\sigma_{ab}(x \vert u) = \theta(u-x)\langle \sigma \rangle_{a}  + \theta(x-u)\langle \sigma \rangle_{b} 
\label{sharp}
\EN
the magnetization at a point $x$ on this line for the given interface configuration, the average magnetization
\EQ
\langle \sigma(x,y) \rangle_{ab}^\textrm{sharp} = \int_{\mathbb{R}} \rd u \, \sigma_{ab}(x \vert u) \, p(u,y)
\EN
coincides with (\ref{single17}) for\footnote{The Gaussian passage probability density (\ref{single19}), with a width shrinking to zero at the boundary condition changing points, gives to the interface the property of a Brownian bridge, which has been rigorously proved for the Ising model \cite{GI} and the Potts model \cite{CIV}.}
\EQ
\label{single19}
p(x,y) = \frac{\textrm{e}^{-\chi^2}}{\sqrt{\pi}\,\kappa\lambda} \,.
\EN
We also see that (\ref{branching}) corresponds to adding to (\ref{sharp}) the local term $(\mathscr{C}_{ab}^\sigma/m)\delta(x-u)$, which represents a deviation from sharp phase separation and is the first manifestation of an internal structure of the interface. A typical effect contributing to (\ref{branching}) is the bifurcation and recombination of the interface depicted in Fig.~\ref{configurations}b; we see from the factor of $\lambda$ in the denominator of (\ref{branching}) that it is suppressed as $R^{-1/2}$. The formation of such bubbles requires three different phases, and the term (\ref{branching}) is indeed absent in the Ising model, in which the magnetization is odd in $x$ by symmetry. The first branching effect in the Ising model (trifurcation, Fig.~\ref{configurations}c) contributes to the subsequent term of the low energy expansion and is suppressed\footnote{Within the saddle point evaluation of (\ref{single08}) at large $R$ each additional power of rapidity in the product $f_{ab}^*(\theta)f_{ab}(\theta')\mathcal{M}_{ab}^\sigma$ contributes a factor $R^{-1/2}$.} as $R^{-1}$ at large $R$.

\begin{figure}[htbp]
\centering
\includegraphics[width=2.5cm]{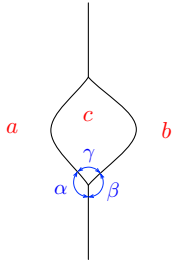}
\caption{A bubble of phase $c$ contributing to the internal structure of the interface between phases $a$ and $b$.}
\label{bubble}
\end{figure}

The bifurcation of the interface requires the presence in the theory of a three-kink vertex, corresponding to the bound state formation $|K_{ac}(\theta_1)K_{cb}(\theta_2)\rangle\to|K_{ab}(0)\rangle$ for some resonant value $\theta_1-\theta_2=i\gamma$, $\gamma\in(0,\pi)$. Relativistic kinematics yields the relation $m_{ab}^2=m_{ac}^2+m_{cb}^2+2m_{ac}m_{cb}\cos\gamma$ among the masses of the three kinks. In view of (\ref{tension}) this becomes the well known relation among the components of the superficial tensions at each vertex of the bubble \cite{Dietrich} (Fig.~\ref{bubble}). It also follows from (\ref{rapidity}) that energy conservation at the vertex becomes the equilibrium condition
\EQ
\Sigma_{ab}+\Sigma_{ac}\cos\alpha+\Sigma_{cb}\cos\beta=0\,.
\EN

\section{Double interfaces and intermediate phases}
\begin{figure}[htbp]
\centering
\includegraphics[width=5.5cm]{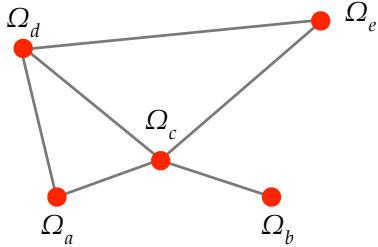}
\caption{A vacuum structure including non-adjacent vacua.}
\label{vacuastructure}
\end{figure}

We now consider the case in which, still starting with the $ab$ boundary conditions of Fig.~\ref{rectangle}, phases $a$ and $b$ are not adjacent. More precisely, we consider the simplest case of this type, the one in which the minimal path between $|\Omega_{a}\rangle$ and $|\Omega_{b}\rangle$ is a two-kink state $|K_{ac}K_{cb}\rangle$ passing through a third vacuum $|\Omega_{c}\rangle$ (an example of this kind is shown in Fig.~\ref{vacuastructure}). 
This means that now the expansion (\ref{single01}) of the boundary state is replaced by
\EQ
\label{double02}
\vert B_{ab}(x_{0};t) \rangle = \textrm{e}^{-itH+ix_{0}P} \, \Biggl[ \, \sum_{c \neq a,b} \int_{\mathbb{R}^{2}} \frac{\rd \theta_{1}}{2\pi} \frac{\rd \theta_{2}}{2\pi} \, f_{acb}(\theta_{1},\theta_{2}) \, \vert K_{ac}(\theta_{1}) K_{cb}(\theta_{2}) \rangle +\dots \Biggr],
\EN
where the summation over $c$ indicates that, in general, there can be more than one two-kink path interpolating between $|\Omega_{a}\rangle$ and $|\Omega_{b}\rangle$. For simplicity we refer to the case in which the lightest state $|K_{ac}K_{cb}\rangle$, is made of two kinks with the same mass $m$. Then we can stipulate that the sum in (\ref{double02}) includes only the states $|K_{ac}K_{cb}\rangle$ with mass $2m$, with the dots including all heavier states and contributing subleading terms in the large $R$ expansion. Plugging (\ref{double02}) into (\ref{single02}) then gives
\EQ
\label{double04}
\mathcal{Z}_{ab}(R)\simeq \sum_{c,d\neq a,b} \int_{\mathbb{R}^{4}} \frac{\rd\theta_{1} \rd\theta_{2} \rd\theta_{3} \rd\theta_{4}}{(2\pi)^{4}} \, \mathcal{F}_{ab,cd} \, \mathcal{M}_{ab,cd} \, \mathcal{Y},
\EN
where we defined
\bea \nonumber
\mathcal{F}_{ab,cd}(\theta_{1},\theta_{2},\theta_{3},\theta_{4}) & \equiv & f_{acb}(\theta_{1},\theta_{2}) f_{adb}^{*}(\theta_{3},\theta_{4}) \\ \label{Mabcd}
\mathcal{M}_{ab,cd}(\theta_{1},\theta_{2}\vert\theta_{3},\theta_{4}) & \equiv & \langle K_{bd}(\theta_{3}) K_{da}(\theta_{4}) \vert K_{ac}(\theta_{1}) K_{cb}(\theta_{2}) \rangle \\ \nonumber
\mathcal{Y}(\theta_{1},\theta_{2},\theta_{3},\theta_{4}) & \equiv & Y^{-}(\theta_{1})Y^{-}(\theta_{2})Y^{+}(\theta_{3})Y^{+}(\theta_{4}) \\
Y^{\pm}(\theta) & \equiv & \textrm{e}^{m\bigl[ -\frac{R}{2}\cosh\theta \pm i x \sinh\theta \bigr]}\,.
\eea

\begin{figure}[htbp]
\centering
\includegraphics[width=3.5cm]{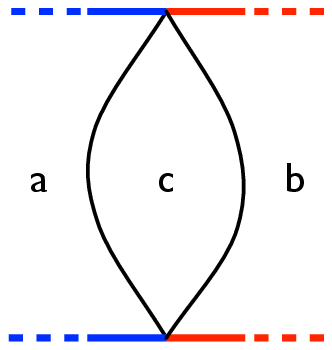}\hspace{1cm}
\includegraphics[width=3.5cm]{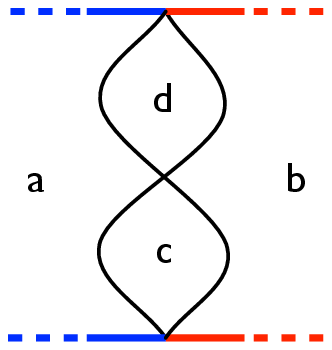}
\caption{The two contributions to (\ref{double09}).}
\label{Mdiagrams}
\end{figure}

In our framework, the two-kink states are asymptotic states that can be either incoming or outgoing, the two basis being related by the scattering operator. Since the large $R$ limit we are interested in projects towards small rapidities, scattering processes take place at energies below any particle production threshold, and are then elastic. In particular, two-kink states scatter into two-kink states, and we can write
\EQ
|K_{ac}(\theta_{1})K_{cb}(\theta_{2})\rangle=\sum_d S_{ab}^{cd}(\theta_{1}-\theta_{2}) \,|K_{ad}(\theta_{2})K_{db}(\theta_{1})\rangle\,,
\label{FZ}
\EN
where
\EQ
S_{ab}^{cd}(\theta_1-\theta_2) = \fig{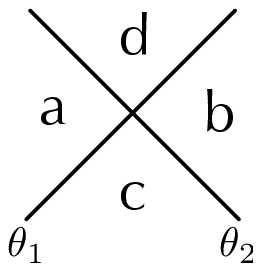}
\label{amplitudes}
\EN
are the two-kink scattering amplitudes, in which all kinks have mass $m$ and initial and final rapidities coincide by two-dimensional energy-momentum conservation; in (\ref{FZ}) we also stipulated that $\theta_1>\theta_2$ and that kinks are ordered according to decreasing (resp. increasing) rapidity for incoming (resp. outgoing) states. The unitarity condition associated to (\ref{FZ}) then reads
\EQ
\sum_{e} S_{ab}^{ce}(\theta)S_{ab}^{ed}(-\theta)=\delta_{cd}\,.
\label{unitarity}
\EN
Since the large $R$ limit leads to consider rapidities which tend to zero, the essential information we need from the scattering theory is the threshold value $S_{ab}^{cd}(0)$ of the amplitudes. The models to which we will specialize in the next sections satisfy
\EQ
S_{ab}^{cd}(0)=-\delta_{cd}\,,
\label{s0}
\EN
and this is the case that we consider in the following\footnote{It is possible that (\ref{s0}) is a necessary condition for the formation of an intermediate phase.}. The use of (\ref{s0}) into (\ref{FZ}) with $\theta_1=\theta_2$ shows that the states $|K_{ac}(\theta)K_{cb}(\theta)\rangle$ are not allowed. It follows in particular that the amplitudes $f_{acb}$ in (\ref{double02}) need to vanish when $\theta_1=\theta_2$, and can be written as
$f_{acb}(\theta_{1},\theta_{2})\simeq c_{acb} \, \theta_{12}$ at small rapidities, where we defined $\theta_{ij}\equiv\theta_i-\theta_j$. As a consequence
\EQ
\mathcal{F}_{ab,cd}(\theta_{1},\theta_{2},\theta_{3},\theta_{4})\simeq c_{acb}c^*_{adb}\,\theta_{12}\theta_{34}
\label{Fabcd}
\EN
at small rapidities. Concerning the product (\ref{Mabcd}), it is the sum of the two terms\footnote{Due to (\ref{unitarity}) terms with more than one bulk crossing reduce to those of Fig.~\ref{Mdiagrams}.} depicted in Fig.~\ref{Mdiagrams}, and reads
\EQ
\label{double09}
\mathcal{M}_{ab,cd}(\theta_{1},\theta_{2}\vert\theta_{3},\theta_{4})=(2\pi)^{2}[\delta(\theta_{14})\delta(\theta_{23}) \delta_{cd}+\delta(\theta_{13})\delta(\theta_{24}) \, S_{ab}^{cd}(\theta_{12})]\,.
\EN

With (\ref{Fabcd}) and (\ref{double09}) we can proceed to the saddle point evaluation of (\ref{double04}), obtaining
\EQ
\label{part2}
\mathcal{Z}_{ab}(R)\simeq \zeta_{ab}\int_{\mathbb{R}^{2}} \frac{\rd \theta_{1}\rd \theta_{2}}{(2\pi)^{2}} \, \theta_{12}^{2} \, \textrm{e}^{-mR(\cosh\theta_{1}+\cosh\theta_{2})} \simeq \zeta_{ab}\frac{\textrm{e}^{-2mR}}{\pi(mR)^{2}}\,,
\EN
with
\EQ
\zeta_{ab} = \sum_{c,d\neq a,b} c_{acb}c_{adb}^{*} \Bigl[ S_{ab}^{cd}(0) - \delta_{cd} \Bigr] = -2\sum_{c\neq a,b} |c_{acb}|^{2}.
\EN
The interfacial tension (\ref{single04}) is now $2m$, as expected for the double interface of Fig.~\ref{configurations}d.

The magnetization (\ref{single05}) with the boundary state (\ref{double02}) becomes
\bea \nonumber
\langle \sigma(x,y) \rangle_{ab} & \simeq & \frac{1}{\mathcal{Z}_{ab}(R)} \sum_{c,d \neq a,b} \int_{\mathbb{R}^{4}} \frac{\rd \theta_{1} \rd \theta_{2} \rd \theta_{3} \rd \theta_{4}}{(2\pi)^4} \, \mathcal{F}_{ab,cd}(\theta_{1},\theta_{2},\theta_{3},\theta_{4}) \times \\ \nonumber
& \times & \langle K_{bd}(\theta_{3}) K_{da}(\theta_{4}) \vert \textrm{e}^{-\frac{R}{2}H} \, \overbrace{\textrm{e}^{ixP+yH} \, \sigma(0,0) \, \textrm{e}^{-ixP-yH}}^{\sigma(x,y)} \, \textrm{e}^{-\frac{R}{2}H} \vert K_{ac}(\theta_{1}) K_{cb}(\theta_{2}) \rangle \\ \nonumber
& = &\frac{1}{\mathcal{Z}_{ab}(R)} \sum_{c,d \neq a,b} \int_{\mathbb{R}^{4}} \frac{\rd \theta_{1} \rd \theta_{2} \rd \theta_{3} \rd \theta_{4}}{(2\pi)^4} \, \mathcal{F}_{ab,cd}(\theta_{1},\theta_{2},\theta_{3},\theta_{4}) \, \mathcal{Y}^{\star}(\theta_{1},\theta_{2},\theta_{3},\theta_{4}) \\
& \times & \, \mathcal{M}_{ab,cd}^{\sigma}(\theta_{1},\theta_{2}\vert\theta_{3},\theta_{4}),
\label{double10}
\eea
where 
\EQ
\mathcal{Y}^{\star}(\theta_{1},\dots,\theta_{4}) \equiv \mathcal{Y}(\theta_{1},\dots,\theta_{4}) \, \textrm{e}^{my(\cosh\theta_{3}+\cosh\theta_{4}-\cosh\theta_{1}-\cosh\theta_{2}) }\,,
\label{ystar}
\EN
\EQ
\mathcal{M}_{ab,cd}^{\sigma}(\theta_{1},\theta_{2}\vert\theta_{3},\theta_{4}) \equiv \langle K_{bd}(\theta_{3}) K_{da}(\theta_{4}) \vert \sigma(0,0) \vert K_{ac}(\theta_{1}) K_{cb}(\theta_{2}) \rangle\,.
\label{sigma4}
\EN
Analogously to what discussed in the previous section for the two-leg case, the matrix element (\ref{sigma4}) contains a connected part, that we will denote $\mathcal{M}^{\sigma,\textrm{conn}}_{ab,cd}$, and a number of disconnected contributions; pictorially
\bea
\label{double12}
\newfig{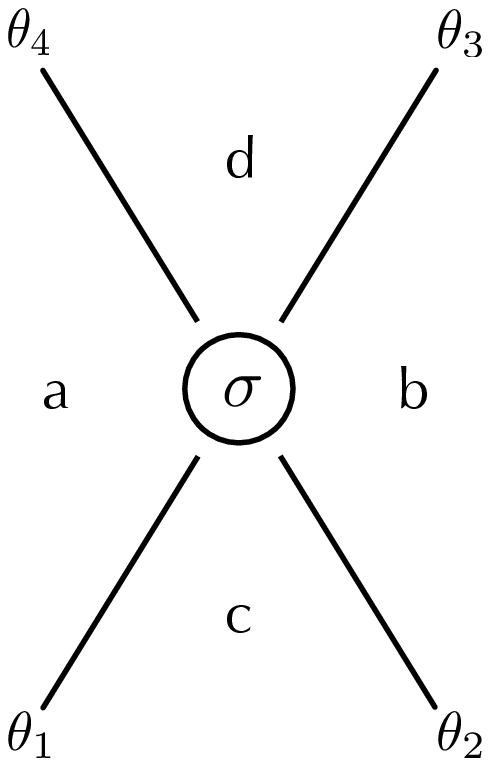} \quad & = & \newfig{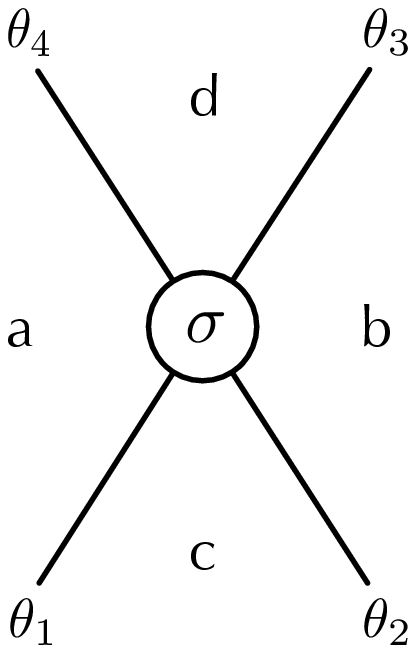} \qquad + \quad \textrm{disconnected parts}\,.
\eea
As in the two-leg case, the possibility of performing the decomposition in two different ways, depending on whether the disconnected trajectories pass to the right or to the left of the insertion point of the magnetization operator, leads to kinematical singularities in the connected parts. The residues on these poles are given by the generalization of (\ref{residue}), namely
\bea \nonumber
-i \, \textrm{Res}_{\theta_{1} = \theta_{3}} \, \newfig{m1} \qquad & = & \qquad \newfig{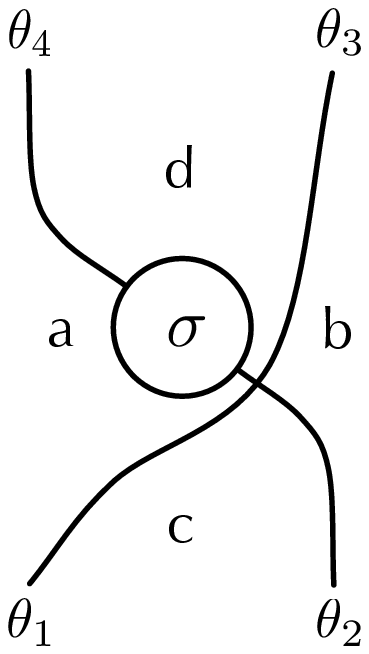} \quad - \quad \newfig{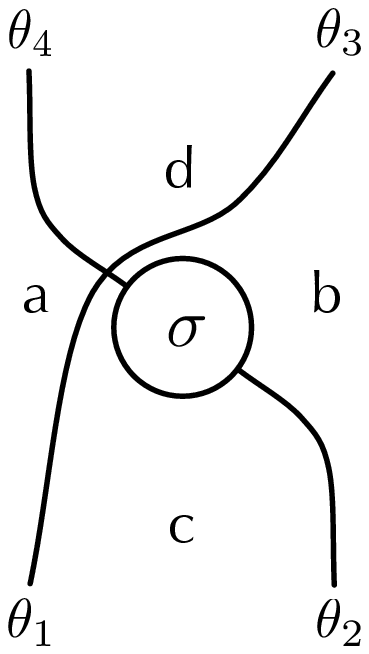} \\ \nonumber
& = & S_{ab}^{cd}(0) \newfig{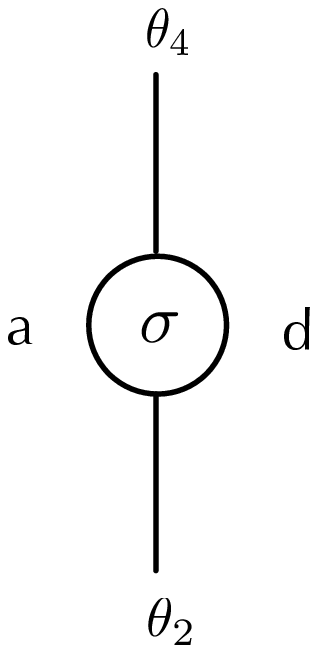} - \quad S_{ab}^{cd}(0) \newfig{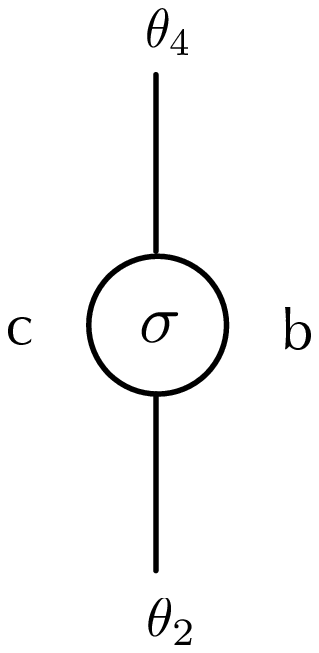} \\ 
& \simeq & \frac{i S_{ab}^{cd}(0)}{\theta_{24}} \biggl[ \langle \sigma \rangle_{a} - \langle \sigma \rangle_{d} - \langle \sigma \rangle_{c} + \langle \sigma \rangle_{b} \biggr],
\eea
where we work directly in the limit $\theta_1,\ldots\theta_4\to 0$ and used (\ref{pole}) in the last line (we do not need to keep track of the $i\epsilon$ prescriptions here). Similarly,
\bea \nonumber
-i \, \textrm{Res}_{\theta_{1}=\theta_{4}} \, \newfig{m1} \quad & = & \quad \newfig{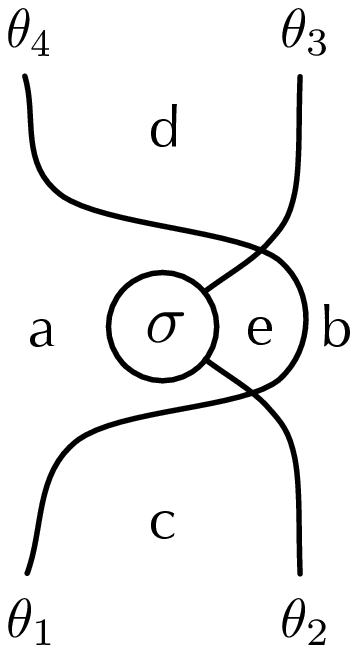} \quad - \quad \delta_{cd} \, \newfig{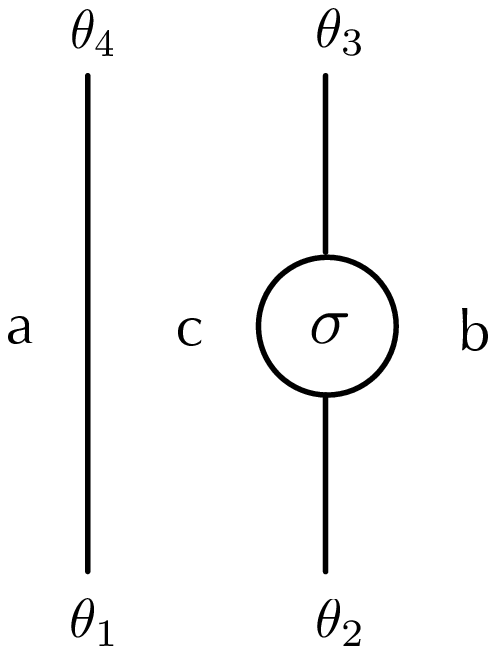} \\ \nonumber
& = & \sum_{e} S_{ab}^{ce}(0)S_{ab}^{ed}(0) \newfig{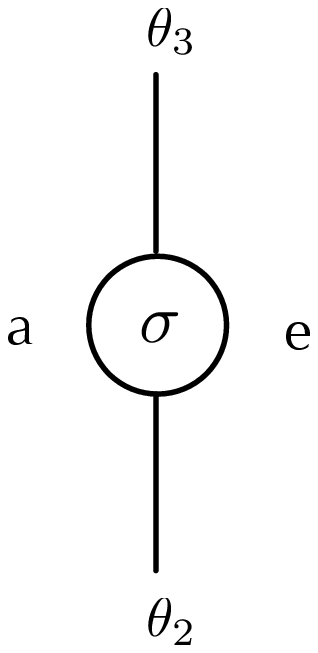}\,\,\, - \,\,\,\delta_{cd} \newfig{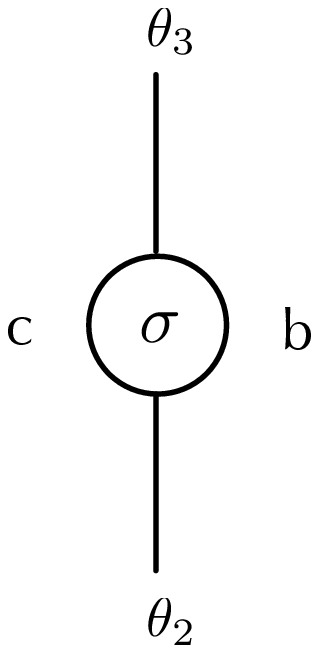} \\ 
\nonumber
& \simeq & \frac{i}{\theta_{23}} \biggl[ \sum_{e} S_{ab}^{ce}(0)S_{ab}^{ed}(0) \Bigl[ \langle \sigma \rangle_{a} - \langle \sigma \rangle_{e} \Bigr] - \delta_{cd}\Bigl[ \langle \sigma \rangle_{c} - \langle \sigma \rangle_{b} \Bigr] \biggr] \\ \nonumber
& = & \frac{i}{\theta_{23}} \biggl[ -\sum_{e} S_{ab}^{ce}(0)S_{ab}^{ed}(0) \langle \sigma \rangle_{e} - \delta_{cd}\Bigl[ \langle \sigma \rangle_{c} - \langle \sigma \rangle_{a} - \langle \sigma \rangle_{b} \Bigr] \biggr]\,;
\eea
(\ref{unitarity}) was used in the last line. Analogous results are obtained when the kink with rapidity $\theta_2$ is disconnected, and we have
\bea \nonumber
- \, \textrm{Res}_{\theta_{1}=\theta_{3}} \mathcal{M}_{ab,cd}^{\sigma,\textrm{conn}}(\theta_{1},\dots,\theta_{4}) & = & \frac{ \mathcal{A}_{ab,cd}}{\theta_{24}}, \\ \nonumber
- \, \textrm{Res}_{\theta_{1}=\theta_{4}} \mathcal{M}_{ab,cd}^{\sigma,\textrm{conn}}(\theta_{1},\dots,\theta_{4}) & = & \frac{ \mathcal{B}_{ab,cd}}{\theta_{23}}, \\ \nonumber
- \, \textrm{Res}_{\theta_{2}=\theta_{4}} \mathcal{M}_{ab,cd}^{\sigma,\textrm{conn}}(\theta_{1},\dots,\theta_{4}) & = & \frac{ \mathcal{A}_{ab,cd}}{\theta_{13}}, \\
-\, \textrm{Res}_{\theta_{2}=\theta_{3}} \mathcal{M}_{ab,cd}^{\sigma,\textrm{conn}}(\theta_{1},\dots,\theta_{4}) & = & \frac{ \mathcal{B}_{ab,cd}}{\theta_{14}},\nonumber
\eea
with
\bea \nonumber
\mathcal{A}_{ab,cd} & = & S_{ab}^{cd}(0)[ \langle \sigma \rangle_{a} + \langle \sigma \rangle_{b} - \langle \sigma \rangle_{c} - \langle \sigma \rangle_{d}], \\
\mathcal{B}_{ab,cd} & = & \delta_{cd} \bigl[ \langle \sigma \rangle_{a} + \langle \sigma \rangle_{b} - \langle \sigma \rangle_{c} \bigr] - \sum_{e} S_{ab}^{ce}(0)S_{ab}^{ed}(0)\langle \sigma \rangle_{e}\,.\nonumber
\eea
The condition (\ref{s0}) simplifies the result to $\mathcal{A}_{ab,cd} = \delta_{cd}(2\langle \sigma \rangle_{c} - \langle \sigma \rangle_{a} - \langle \sigma \rangle_{b})= - \mathcal{B}_{ab,cd}$, and we obtain
\bea
\label{double14}
\mathcal{M}_{ab,cc}^{\sigma,\textrm{conn}}(\theta_{1},\theta_{2},\theta_{3},\theta_{4})& \simeq &[2\langle \sigma \rangle_{c} - \langle \sigma \rangle_{a} - \langle \sigma \rangle_{b}]\,\frac{\theta_{12}\theta_{34}}{\theta_{13}\theta_{14}\theta_{23}\theta_{24}}\,,\\
\mathcal{M}_{ab,cd}^{\sigma,\textrm{conn}}(\theta_{1},\theta_{2},\theta_{3},\theta_{4}) &\simeq &\mathscr{C}^\sigma_{ab,cd}\,\theta_{12} \theta_{34}\,,\hspace{1cm}c\neq d\,,
\label{double14a}
\eea
at small rapidities. As in (\ref{Fabcd}), the prefactor $\theta_{12} \theta_{34}$ accounts for the property (\ref{s0}), and provides the leading term for $c\neq d$, when all the residues vanish. The value of the constant $\mathscr{C}^\sigma_{ab,cd}$ depends on the form of the scattering amplitudes $S_{ab}^{cd}(\theta)$ for $\theta\neq 0$. Notice, however, that the total degree of (\ref{double14a}) in the rapidity variables exceeds by four units that of (\ref{double14}), so that the contribution of (\ref{double14a}) to the magnetization is subleading\footnote{Within the saddle point evaluation of (\ref{double10}) at large $R$ each additional power of rapidity in the product $\mathcal{F}_{ab,cd}\mathcal{M}_{ab,cd}^\sigma$ contributes a factor $1/\sqrt{R}$.} at large $R$ with respect to the leading as well as to some of the subleading terms we are omitting in (\ref{double14}). This means that (\ref{double14a}) must be ignored at this level of the calculation. Hence the contribution of the connected parts is
\bea
\langle \sigma(x,y) \rangle_{ab}^{\textrm{conn}}&\simeq &\frac{1}{\mathcal{Z}_{ab}(R)} \sum_{c\neq a,b} \int_{\mathbb{R}^{4}} \frac{\rd \theta_{1} \rd \theta_{2} \rd \theta_{3} \rd \theta_{4}}{(2\pi)^{4}} \, \mathcal{F}_{ab,cd} \mathcal{Y}^{\star} \, \mathcal{M}_{ab,cc}^{\sigma,\textrm{conn}}\nonumber \\
&\simeq &\frac{\mathcal{G}(\chi)-1}{4\sum_{c\neq a,b}|c_{acb}|^{2}}\sum_{c\neq a,b}|c_{acb}|^{2} [\langle \sigma \rangle_{a}+\langle \sigma \rangle_{b}- 2\langle \sigma \rangle_{c}]\,,
\label{conn}
\eea
with a function $\mathcal{G}(\chi)$ which is computed in the appendix and reads
\EQ
\label{double16}
\mathcal{G}(\chi)=-\frac{2}{\pi} \textrm{e}^{-2\chi^{2}} - \frac{2}{\sqrt{\pi}} \, \chi \, \textrm{erf}(\chi) \, \textrm{e}^{-\chi^{2}} + \textrm{erf}^{2}(\chi)\,.
\EN

Consider now the disconnected parts of (\ref{double12}). We ignore those with two disconnected trajectories, since they contribute to the magnetization (\ref{double10}) only an additive constant that we will fix anyway from the condition $\langle\sigma(+\infty,y)\rangle_{ab}=\langle\sigma\rangle_b$. For the contributions with a single disconneted trajectory we use the notations
\EQ \nonumber
\mathcal{D}_{1423}^{(L)} = \newfig{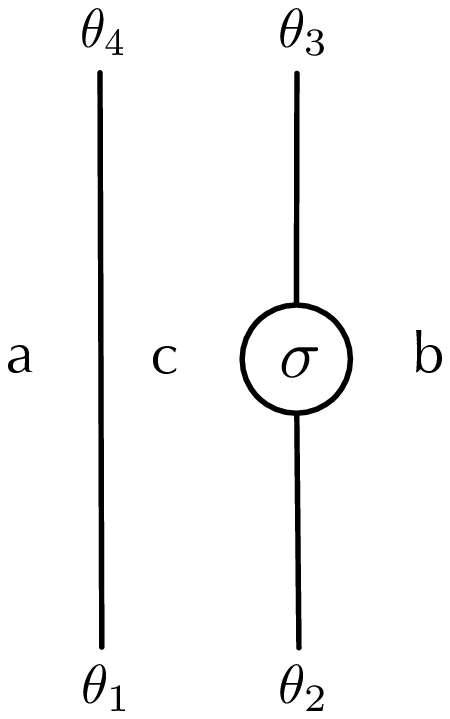} \quad,
\quad \mathcal{D}_{2314}^{(L)} = \sum_e\newfig{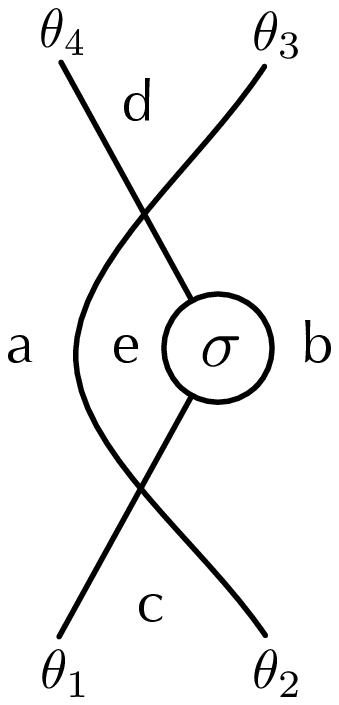},
\quad \mathcal{D}_{1324}^{(L)} = \newfig{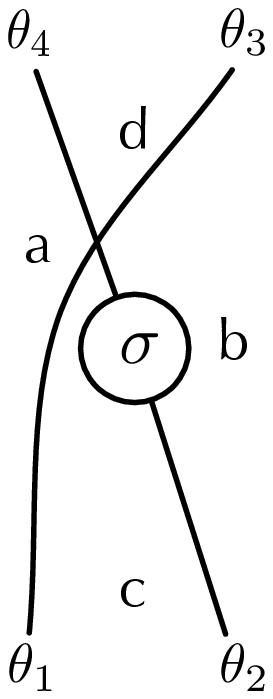} ,
\, \quad \mathcal{D}_{2413}^{(L)} = \newfig{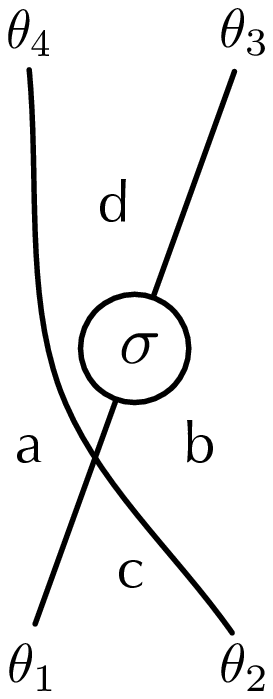},
\EN
and
\EQ \nonumber
\mathcal{D}_{1423}^{(R)} = \sum_e\newfig{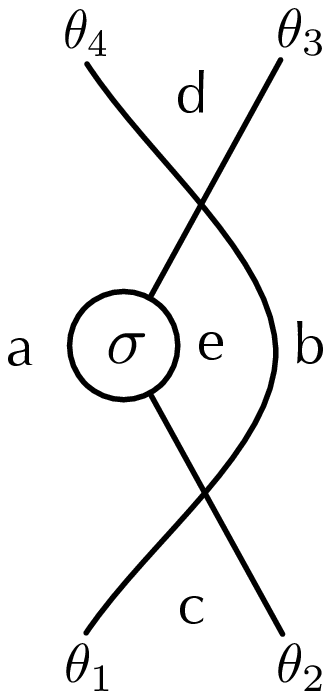} \quad,
\,\, \mathcal{D}_{2314}^{(R)} = \newfig{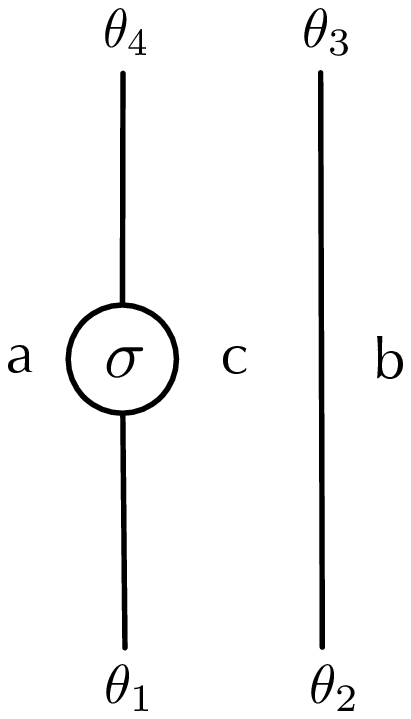} \quad ,
\quad \mathcal{D}_{1324}^{(R)} = \newfig{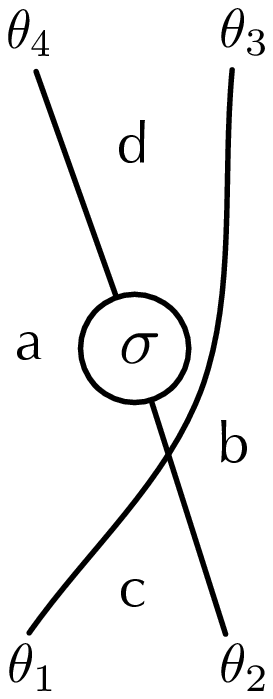},
\,\, \mathcal{D}_{2413}^{(R)} = \newfig{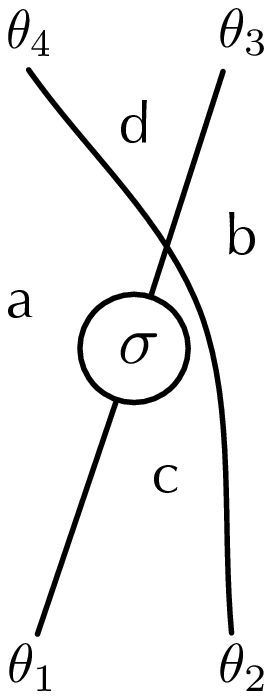},
\EN
depending on the left or right passage prescription. In the limit $\theta_1,\ldots,\theta_4\to 0$, taking into account (\ref{s0}) and (\ref{residue}), we have\footnote{It is understood that the indices $ijkl$ take only the four combinations given above.}
\bea
\mathcal{D}_{ijkl}^{(L)} &\simeq & (-1)^{i+j}2\pi \delta_{cd}\,\delta(\theta_{ij})\,i\frac{\langle \sigma \rangle_{c} - \langle \sigma \rangle_{b}}{\theta_{kl}}\,,\\
\mathcal{D}_{ijkl}^{(R)} &\simeq & (-1)^{i+j}2\pi \delta_{cd}\,\delta(\theta_{ij})\,i\frac{\langle \sigma \rangle_{a} - \langle \sigma \rangle_{c}}{\theta_{kl}}\,,
\eea
from which we see that the two prescriptions are inequivalent. The natural idea to take the average 
\EQ
\mathcal{D}_{ijkl}\equiv\frac{\mathcal{D}_{ijkl}^{(L)}+\mathcal{D}_{ijkl}^{(R)}}{2}=(-1)^{i+j}\pi \delta_{cd}\,\delta(\theta_{ij})\,i\frac{\langle \sigma \rangle_{a} - \langle \sigma \rangle_{b}}{\theta_{kl}}\,
\label{average}
\EN
is the right one. Indeed, as seen in the previous section, single pole terms of this type generate a difference between the values of the magnetization at $x=-\infty$ and $x=+\infty$ proportional to the residue on the pole; we are going to see that (\ref{average}) produces precisely the required difference\footnote{The connected part is even in $x$ and does not contribute to the difference between the asymptotic values.} $\langle \sigma \rangle_{a} - \langle \sigma \rangle_{b}$. The contribution of $\mathcal{D}_{ijkl}$ to the magnetization (\ref{double10}) is $(\langle \sigma \rangle_{a} - \langle \sigma \rangle_{b})/2$ times
\bea
\nonumber
\Delta_{ijkl}(x,y) &=& (-1)^{i+j}\sum_{c\neq a,b}\frac{|c_{acb}|^{2}}{\mathcal{Z}_{ab}(R)} \int_{\mathbb{R}^{4}} \frac{\rd\theta_{1}\rd\theta_{2}\rd\theta_{3}\rd\theta_{4}}{(2\pi)^{4}} \, \theta_{12}\theta_{34} \, \frac{2\pi i \, \delta(\theta_{ij})}{\theta_{kl}} \, \mathcal{Y}^{\star}(\theta_{1},\theta_{2},\theta_{3},\theta_{4})\\
&=&\frac{1}{4}{\cal L}(\chi)+\mbox{constant}\,;
\label{double18}
\eea
the integral is performed in the appendix and gives
\EQ
\mathcal{L}(\chi) = \frac{\chi}{\sqrt{\pi}} \, \textrm{e}^{-\chi^{2}}  - \textrm{erf}(\chi)\,.
\label{L}
\EN
Recalling that there are four choices of $ijkl$, we obtain
\EQ
\label{double19}
\langle \sigma(x,y) \rangle_{ab}^{\textrm{disc}} \simeq \frac{\langle \sigma \rangle_{a} - \langle \sigma \rangle_{b}}{2} \mathcal{L}(\chi) + \mbox{constant}
\EN
for the contribution of the disconnected parts to the magnetization. With this we are ready to apply the findings of this section to specific models.

\section{$q$-state Potts model}
\begin{figure}[htbp]
\centering
\includegraphics[width=9cm]{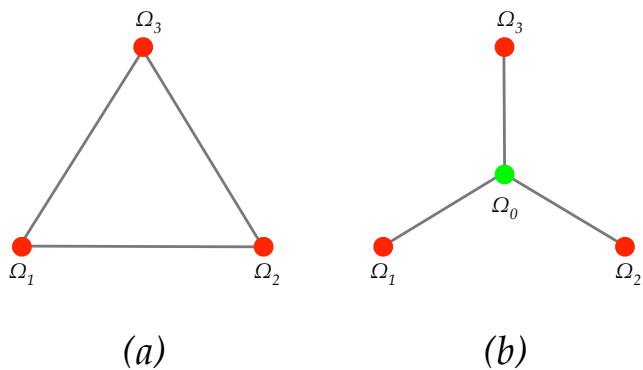}
\caption{Vacuum and kink structure of the three-state Potts model at first order transition points in the pure case (a) and in the dilute case (b).}
\label{dvacua}
\end{figure}

The $q$-state Potts model is a generalization of the Ising model to the case in which the spin variable takes $q$ values (colors), and is characterized by the invariance of the Hamiltonian under global permutations of the colors \cite{Wu}. For ferromagnetic interaction in two dimensions it undergoes at a critical temperature $T_c$ a phase transition which is continuous for $q\leq 4$ and first order for $q>4$ \cite{Baxter} (see \cite{paraf} for a derivation in the continuum). Hence, for $q\leq 4$ and $T<T_c$ there is a scaling limit corresponding to a field theory with $q$ degenerate vacua and kinks interpolating between each pair of them (see Fig.~\ref{dvacua}.a). Phase separation is necessarily of the type discussed in section~2 and was analyzed in detail in \cite{DV}.

When annealed vacancies are introduced, the transition for $q<4$ stays continuous up to a critical value $\rho_c$ of the vacancy density, above which it becomes first order. As $q$ is varied, there is then a tricritical line for $T=T_c(\rho)$, $\rho=\rho_c$, which is known to coalesce with the critical line of the undilute ($\rho=0$) model at $q=4$ \cite{Wu}. On the first order surface $T=T_c(\rho)$, $\rho>\rho_c$ the ferromagnetic vacua $|\Omega_i\rangle$, $i=1,\ldots,q$,  are degenerate with the disordered one $|\Omega_0\rangle$. The elementary excitations are the kinks $|K_{0i}\rangle$ running between the disordered vacuum and the ferromagnetic ones (Fig.~\ref{dvacua}.b), as two ferromagnetic vacua will be related by $|K_{i0}K_{0j}\rangle$. In principle such two-kink configurations could give rise to stable bound state kinks $|K_{ij}\rangle$, that would make the vacua $|\Omega_i\rangle$ and $|\Omega_j\rangle$ adjacent. This, however, is not the case. Indeed, the field theory corresponding to the scaling limit on the first order surface is integrable\footnote{The scaling limit without dilution is also integrable \cite{CZ}.}, and the spectrum of excitations and the scattering amplitudes are known exactly \cite{dilute}. The kinks $|K_{0i}\rangle$ and $|K_{i0}\rangle$ do not form bound states and are the only single-particle excitations of the theory. As a consequence the vacua $|\Omega_i\rangle$ and $|\Omega_j\rangle$ are not adjacent and the strip with boundary conditions $i$ on the left and $j$ on the right will give rise to a double interface containing a bubble of the disordered phase. We now apply to this case the formalism of the previous section.

\begin{figure}[htbp]
\centering
\includegraphics[width=10cm]{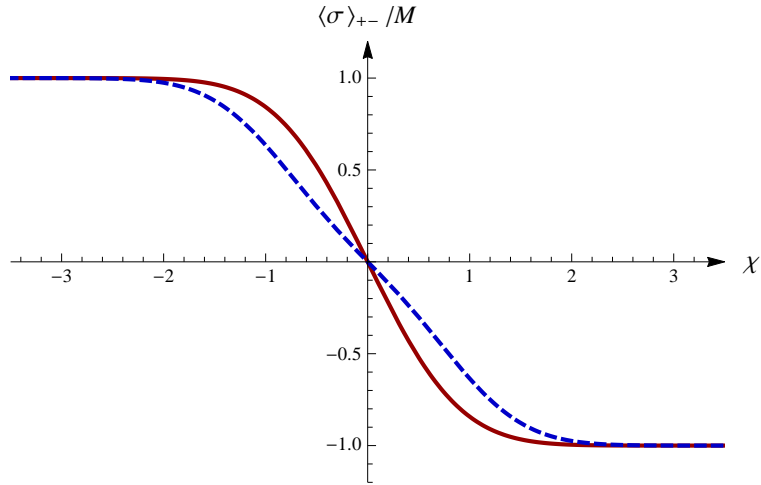}
\caption{Ising magnetization profile $\langle\sigma\rangle_{+-}/M$ at the first order transition in the pure model (continuous curve), and in the dilute model (dashed curve). The presence of an intermediate disordered phase in the dilute case flattens the profile.}
\label{ising}
\end{figure}

\begin{figure}[htbp]
\centering
\includegraphics[width=12cm]{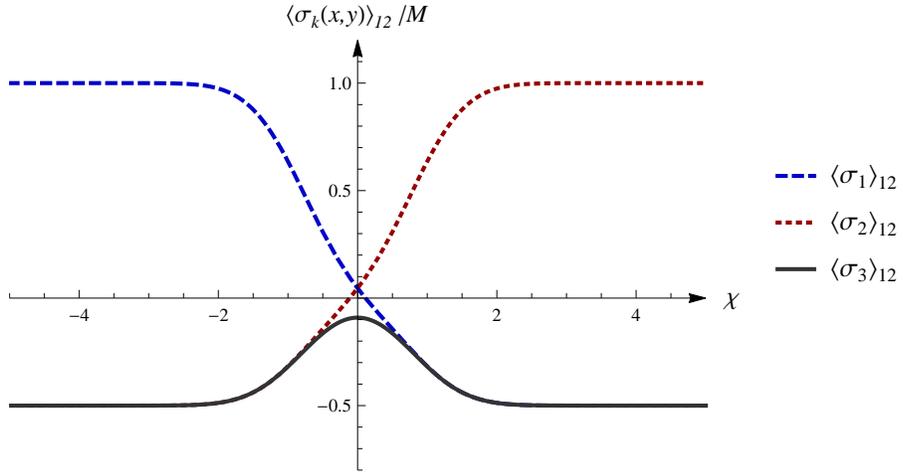}
\caption{Magnetization profiles (\ref{dil2}) for $q=3$.}
\label{dilute}
\end{figure}

\begin{figure}[htbp]
\centering
\includegraphics[width=8.2cm]{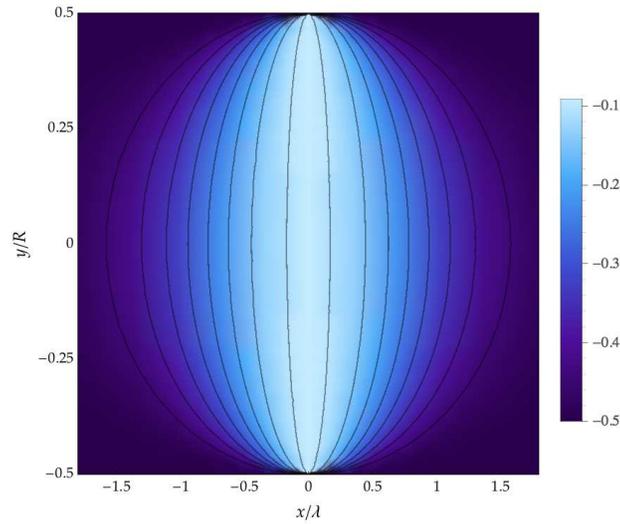}
\caption{The magnetization $\langle \sigma_{3}(x,y) \rangle_{12}/M$ for $q=3$. The curves are the ellipses $\frac{x^{2}}{\lambda^{2}C^{2}} + \frac{4y^{2}}{R^{2}} = 1$, corresponding to $\chi=C$, and then to constant magnetization.}
\label{ising_2}
\end{figure}

All kinks have the same mass as a consequence of permutational symmetry of the colors, which is unaffected by dilution. Since the only two-kink state connecting two different ferromagnetic vacua $|\Omega_i\rangle$ and $|\Omega_j\rangle$ is $|K_{i0}K_{0j}\rangle$, the intermediate index $c$ in (\ref{double02}) is fixed to the value 0. Similarly, 
\EQ
S_{ij}^{cd}(\theta) = \delta_{c0}\delta_{d0} \, S_{ij}^{00}(\theta);
\EN
in addition $S_{ij}^{00}(0)=-1$ \cite{dilute}, so that (\ref{s0}) is fulfilled.
If $s(x,y)$ is the color of a spin at site $(x,y)$, we define the spin variables
\EQ
\sigma_{k}(x,y) = \delta_{k,s(x,y)}-\frac{1}{q}\,,\hspace{2cm}k=1,\ldots,q\,,
\EN
and use the same notation $\sigma_{k}(x,y)$ for the corresponding components of the magnetization operator in the continuum; they satisfy $\sum_{k=1}^q\sigma_k(x,y)=0$. The symmetry gives
\EQ
\label{dil1}
\langle \sigma_{k} \rangle_{j} = \frac{q\delta_{kj}-1}{q-1}M
\EN
in the pure ferromagnetic phases, and
\EQ
\langle \sigma_{k} \rangle_0=0
\label{vev0}
\EN
in the pure disordered phase. Taking all this into account (\ref{conn}) gives
\EQ
\langle \sigma_{k}(x,y) \rangle_{ij}^{\textrm{conn}}\simeq\frac{ \langle \sigma_{k} \rangle_{i} + \langle \sigma_{k} \rangle_{j} }{4} \biggl[\mathcal{G}(\chi)-1 \biggr]\,.
\EN
Adding the disconntected contribution (\ref{double19}) and fixing the additive constant by the condition $\langle\sigma_k(+\infty,y)\rangle_{ij}=\langle\sigma_k\rangle_j$ we finally obtain
\EQ
\label{dil2}
\langle \sigma_{k}(x,y) \rangle_{ij}\simeq \frac{ \langle \sigma_{k} \rangle_{i} + \langle \sigma_{k} \rangle_{j} }{4} \biggl[1+\mathcal{G}(\chi)\biggr] + \frac{ \langle \sigma_{k} \rangle_{i} - \langle \sigma_{k} \rangle_{j} }{2} \mathcal{L}(\chi)\,.
\EN
For $q=2$ this becomes $\langle \sigma_{k}(x,y) \rangle_{ij} = (-1)^{\delta_{kj}}M\,\mathcal{L}(\chi)$, from which we see that, with respect to the pure Ising case (\ref{single17}), the effect of dilution and of the formation of the intermediate wetting phase is the appearance of the first term of (\ref{L}) (see Fig.~\ref{ising}). 
The results (\ref{dil2}) are shown in Figs.~\ref{dilute} and \ref{ising_2} for $q=3$.

As for the case of single interface of section~2, the results (\ref{dil2}) admit a probabilistic interpretation in terms of average over configurations of interfaces sharply separating pure phases. Suppose indeed that two such interfaces intersect at $x=u_1$ and $x=u_2$ the horizontal axis of constant $y$ inside the strip with $ab$ boundary conditions, and that they can contain a single phase $c$ in between them. The magnetization corresponding to such a configuration on the line of constant $y$ can then be written as
\EQ
\label{prob_02}
\sigma_{acb}(x \vert u_{1}, u_{2}) = \sigma_{acb}^{\star}(x \vert u_{1}, u_{2}) \, \theta(u_{2}-u_{1}) + \sigma_{acb}^{\star}(x \vert u_{2}, u_{1}) \, \theta(u_{1}-u_{2}),
\EN
with
\EQ
\label{prob_03}
\sigma_{acb}^{\star}(x \vert u_{1}, u_{2}) = \langle \sigma \rangle_{a} \, \theta(u_{1}-x)+ \langle \sigma \rangle_{b} \, \theta(x-u_{2}) + \langle \sigma \rangle_{c} \, \left( \theta(x-u_{1}) - \theta(x-u_{2}) \right).
\EN
The magnetization is obtained averaging over the interface positions,
\EQ
\label{prob_01}
\langle \sigma(x,y) \rangle_{ab,c}^\textrm{sharp} = \int_{\mathbb{R}^{2}} \rd u_{1} \rd u_{2} \, \sigma_{acb}(x \vert u_{1}, u_{2}) \, p(u_{1},u_{2};y)\,,
\EN
with $p(u_{1},u_{2};y)$ the probability density for intersections at $u_1$ and $u_2$.
It is not difficult to check that, for $\langle \sigma \rangle_{c}=0$, the result (\ref{dil2}) is precisely reproduced by the probability density
\EQ
p(x_{1},x_{2};y)=\left( \frac{x_{1}-x_{2}}{\lambda\kappa} \right)^2\,p(x_1,y)p(x_2,y)
=\frac{(\chi_{1}-\chi_{2})^2}{\pi\lambda^2\kappa^2}\,e^{-(\chi_1^2+\chi_2^2)}\,,
\label{jointp}
\EN
which correctly satisfies $\int_{\mathbb{R}^{2}} \rd u_{1} \rd u_{2} \,p(u_{1},u_{2};y) = 1$. Hence we see that the probability density $p(x_1,y)p(x_2,y)$ for non-interacting interfaces gets corrected by the factor $(\chi_1-\chi_2)^2$, whose origin must be traced back to the property (\ref{s0}). For a generic $\langle \sigma \rangle_{c}$ (\ref{jointp}) gives
\EQ
\label{prob_17}
\langle \sigma(x,y) \rangle_{ab,c}^\textrm{sharp} = \frac{\langle \sigma \rangle_{a} + \langle \sigma \rangle_{b} - 2\langle \sigma \rangle_{c}}{4} \mathcal{G}(\chi) + \frac{\langle \sigma \rangle_{a} - \langle \sigma \rangle_{b}}{2} \mathcal{L}(\chi) + \frac{\langle \sigma \rangle_{a} + \langle \sigma \rangle_{b} + 2\langle \sigma \rangle_{c}}{4}.
\EN

\begin{figure}[htbp]
\centering
\includegraphics[width=10cm]{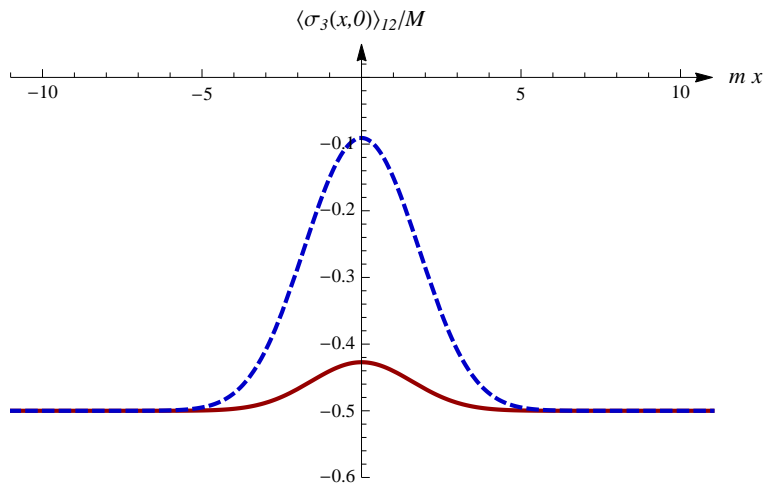}
\caption{Magnetization profile $\langle\sigma_3(x,0)\rangle_{12}/M$ in the three-state Potts model at the first order transition for $mR=10$. For the undilute case (continuous curve) the bump is produced by the branching of the single interface and is suppressed as $R^{-1/2}$. For the dilute case (dashed curve) the bump is due to the disordered intermediate phase and its height persists asymptotically at large $R$.}
\label{sigma312}
\end{figure}

In Fig.~\ref{sigma312} we show $\langle\sigma_3\rangle_{12}$ from (\ref{dil2}) and for the undilute $T<T_c$ three-state Potts model, for which the leading non-constant term is provided by (\ref{branching}) with $\mathscr{C}_{12}^{\sigma_3}=M/(2\sqrt{3})$ \cite{DV}. This latter term accounts for the formation of bubbles of third color depicted in Fig.~\ref{bubble} and corresponding to the vertex $K_{ac}K_{cb}\sim K_{ab}$ which is indeed present in the pure model \cite{CZ}; since all kinks have the same mass, the angles in Fig.~\ref{bubble} are $\alpha=\beta=\gamma=2\pi/3$. Hence Fig.~\ref{sigma312} makes clear the quantitative difference between the effect of the formation of an intermediate disordered phase in the dilute case ($T=T_c$, $\rho>\rho_c$) and that due to the appearance of color 3 via branching and recombination of the single interface in the undilute case. The maximum of (\ref{branching}) decreases as $(mR)^{-1/2}$; for the models we discuss in this and the next section, the correlation length defined by the exponential decay of bulk spin-spin correlations is 
\EQ
\xi=\frac{1}{2m}\,.
\EN

We derived the results (\ref{dil2}), (\ref{jointp}) for the dilute Potts model on the first order surface $T=T_c$, $\rho>\rho_c$, $q<4$. For the undilute model with $q>4$ the phase transition becomes first order and the same vacuum structure considered above ($q$ ferromagnetic vacua degenerate with the disordered vacuum) is present at $T_c$. Strictly speaking, the scaling limit at $T_c$ is only defined in the limit\footnote{It is well known that the Potts model can be continued to real values of $q$ \cite{FK}.} $q\to 4^+$, and the exact solution of the associated field theory was studied in \cite{DCq4}. The vacuum adjacency structure and the property (\ref{s0}) are unchanged with respect to the dilute $q<4$ case, and so is the result for the magnetization profiles. It was also shown in \cite{DCq4} that the field theoretical description remains quantitatively accurate  as long as the correlation length $\xi$ is much larger than lattice spacing. We then expect that (\ref{dil2}) and (\ref{jointp}) are essentially exact in the $q>4$ critical pure model up to values such as $q=10$, where $\xi=10.5$.

It is interesting to notice that the function (\ref{double14}) originally appeared in \cite{McCoyWu} within an exact lattice computation of asymptotics of three-point spin correlators in the Ising model below $T_c$. In the language of this paper, the coincidence is made possible by the fact that the leading non-constant contribution to this asymptotic correlator comes from a two-kink intermediate state. Later on this fact was exploited in \cite{AU} to propose that $\mathcal{G}(\chi)$ gives the magnetization profile across a bubble of down spins surrounded by up spins in the Ising model. It was shown in \cite{KF} by lattice computations that this is indeed the case provided that the pinning points of the interfaces are taken a fixed number of lattice spacings apart on the edges of the strip, and that the configurations with interfaces starting and ending on the same edge are removed by hand, before taking the scaling limit which makes the pinning points on the same edge coalesce. This is a technical way around the basic problem that the Ising model does not possess the three different phases necessary to generate two interfaces with the boundary conditions of Fig.~\ref{rectangle}.

\section{Ashkin-Teller model}
\begin{figure}[htbp]
\centering
\includegraphics[width=5.5cm]{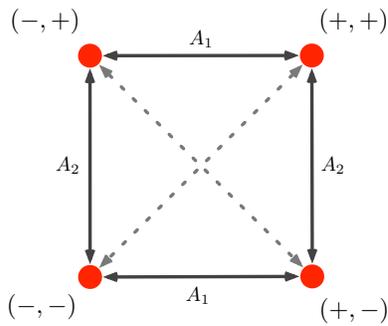}
\caption{Vacuum connectivity in the Ashkin-Teller model. The diagonal kinks are present only for $J_4> 0$.}
\label{at_vacuum}
\end{figure}

The two-dimensional Ashkin-Teller model in defined on the lattice placing at each site $r=(x,y)$ two Ising spins $\sigma_1(r),\sigma_2(r)=\pm 1$, whose interaction is specified by the Hamiltonian
\EQ
\label{at_ham}
\mathcal{H}_{AT} = -\sum_{\langle r_1,r_2 \rangle}\{J[ \sigma_{1}(r_1)\sigma_{1}(r_2) + \sigma_{2}(r_1)\sigma_{2}(r_2)] + J_{4}\,\sigma_{1}(r_1)\sigma_{1}(r_2)\sigma_{2}(r_1)\sigma_{2}(r_2)\} \,,
\EN
where the sum is taken over nearest neighbors; we consider the ferromagnetic case $J>0$. The Hamiltonian is invariant under the exchange 
\EQ
E\,:\,\sigma_1\leftrightarrow\sigma_2\,,
\EN
as well as under separate spin reversals 
\EQ
I_i\,:\,\sigma_i\to -\sigma_i\,.
\EN
The second order phase transition occurring for $J_4=0$, when the two Ising models are decoupled, is known to extend to $J_4\neq 0$ (see e.g. \cite{DG} and references therein). There is then a second order critical line $J_c(J_4)$, and the scaling limit around it is described by the sine-Gordon field theory with Euclidean action
\EQ
\mathcal{A}_{SG}[\varphi] = \int \rd^{2}x \biggl[ \frac{1}{2} \left( \partial \varphi \right)^{2} - \tau \cos \beta\varphi \biggr] \, ,
\EN
where $\tau$ measures the deviation of $J$ from $J_c$, and $\beta$ is the coordinate along the critical line. On the square lattice the relation between $\beta$ and $J_4$ is \cite{Kadanoff}
\EQ
\frac{4\pi}{\beta^2}=1-\frac{2}{\pi}\arcsin\left(\frac{\tanh 2J_4}{\tanh 2J_4-1}\right)\,.
\EN
For $J>J_c$ the spin reversal symmetries are both spontaneously broken and the theory possesses four degenerate vacua $\Omega_{\alpha_1,\alpha_2}$, $\alpha_i=\pm 1$, corresponding to the breaking of $I_i$ in the direction $\alpha_i$. These vacua are connected as shown in Fig.~\ref{at_vacuum} by elementary excitations $A_1$ and $A_2$, which are kinks with respect to $\sigma_1$ and $\sigma_2$, respectively\footnote{Sine-Gordon soliton and anti-soliton correspond to $A_1\pm iA_2$.}, and have the same mass $m$. For $J_4>0$ these kinks form bound states \cite{ZZ} with mass $2m\sin(\pi\beta^2/2(8\pi-\beta^2))$ which run along the diagonals of Fig.~\ref{at_vacuum} and make all vacua adjacent. 
For $J_4\leq 0$, on the other hand, there are no bound states, and the pairs of vacua $\Omega_{\alpha_1,\alpha_2}$ and $\Omega_{-\alpha_1,-\alpha_2}$ are non-adjacent. This is the case we now analyze.

To be definite, in the following we consider $ab$ boundary conditions on the strip with $a=++$ and $b=--$. It follows from the adjacency structure of Fig.~\ref{at_vacuum} that these boundary conditions correspond to a boundary state of the form (\ref{double02}), with $c$ taking the values $+-$ and $-+$; in addition, exchange symmetry implies that the amplitudes $f_{acb}(\theta_1,\theta_2)$ coincide for the two values of $c$. Sine-Gordon field theory is integrable and all the scattering amplitudes are known \cite{ZZ}. For our purposes it is sufficient to know that (see \cite{DG})

\bea
S_{++,--}^{+-,-+}(\theta) &= & S_{++,--}^{-+,+-}(\theta)= \frac{S(\theta)+S_{-}(\theta)}{2} \, , \label{Ssg}\\
S_{++,--}^{+-,+-}(\theta) &= & S_{++,--}^{-+,-+}(\theta)=\frac{S(\theta)-S_{-}(\theta)}{2} \, ,
\eea
where the notation is that of (\ref{amplitudes}), 
\EQ
S_{-}(\theta)=-\frac{\cosh\frac{\pi}{2\xi}(\theta+i\pi)}{\cosh\frac{\pi}{2\xi}(\theta-i\pi)} \, S(\theta) \, ,\hspace{1cm}
\xi = \frac{\pi \beta^{2}}{8\pi-\beta^{2}} \,,
\label{S-}
\EN
and $S(\theta)$ satisfies $S(0)=-1$ for any $\xi$. The decoupling point $J_4=0$ corresponds to $\beta^2=4\pi$, namely to $\xi=\pi$, and $J_4<0$ corresponds to $\xi>\pi$. For $J_4<0$ we have $S_{++,--}^{cd}(0)=-\delta_{cd}$, so that (\ref{s0}) is fulfilled.

We consider the magnetization operators $\sigma_1$, $\sigma_2$ and $\sigma_1\sigma_2$. The symmetries imply that their expectation values in the four pure phases can be written as
\EQ
\langle\sigma_i\rangle_{(\alpha_1,\alpha_2)}=\alpha_i\,M\,,\hspace{1cm}
\langle\sigma_1\sigma_2\rangle_{(\alpha_1,\alpha_2)}=\alpha_1\alpha_2\,\tilde{M}\,.
\label{vevs}
\EN
Concerning the magnetization profiles on the strip with $ab$ boundary conditions, since $c_{acb}$ in (\ref{conn}) does not depend on the allowed values of $c$, (\ref{vevs}) leads to $\langle \sigma_i(x,y) \rangle_{++,--}^{\textrm{conn}}=0$ and $\langle \sigma_{1}\sigma_{2}(x,y) \rangle_{++,--}^{\textrm{conn}}\simeq\tilde{M}(\mathcal{G(\chi)}-1)$\,. Adding the disconnected contribution (\ref{double19}) and fixing the constant at infinty finally gives
\bea
\langle \sigma_i(x,y) \rangle_{++,--}&\simeq & M\,\mathcal{L}(\chi)\,,\label{at1}\\
\langle \sigma_{1}\sigma_{2}(x,y) \rangle_{++,--}&\simeq & \tilde{M}\,\mathcal{G}(\chi)\,.\label{at2}
\eea
It is straightforward to see that these results correspond to the passage probability density (\ref{jointp}) for the two interfaces. Indeed, it is sufficient to sum (\ref{prob_17}) over the two allowed values of $c$, and to consider that each intermediate phase occurs with probability $1/2$.

\begin{figure}[htbp]
\centering
\includegraphics[width=10cm]{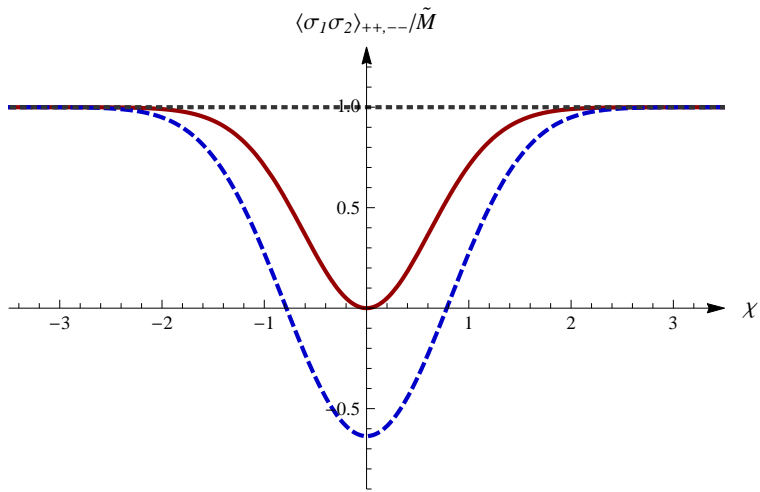}
\caption{Asymptotic magnetization $\langle\sigma_1\sigma_2\rangle_{++,--}/\tilde{M}$ in the Ashkin-Teller model for $J_4<0$ (dashed curve), $J_4=0$ (continuous curve) and $J_4>0$ (dotted curve).}
\label{sigma1sigma2}
\end{figure}
The results (\ref{at1}), (\ref{at2}) have been obtained for $J_4<0$, but do not depend on $J_4$, namely on the interaction between two Ising spins. This corresponds to the fact that the leading large $R$ behavior is entirely determined by the non-crossing condition for the interfaces induced by (\ref{s0}). Consistency then requires that (\ref{s0}) is violated at the decoupling point $J_4=0$, where profiles corresponding to single interfaces must be recovered. Indeed, it follows from (\ref{Ssg})-(\ref{S-}) that precisely at $\xi=\pi$ the threshold value of the amplitudes switches discontinuously to $S_{++,--}^{cd}(0)=\delta_{cd}-1$. A first consequence is that the boundary amplitudes $f_{++,c,--}(0)$ no longer vanish, because the state $|K_{++,c}(\theta)K_{c,--}(\theta)\rangle$ no longer scatters into minus itself; the partition function then becomes the square of (\ref{single02}), as it should. In a similar way, one can adapt to this $J_4=0$ case the rest of the analysis of section~3 and formally recover the single interface results $-M\,\mbox{erf}(\chi)$ instead of (\ref{at1}), and $(M\,\mbox{erf}(\chi))^2$ instead of (\ref{at2}) (Fig.~\ref{sigma1sigma2}). Hence, we see how an arbitrarily small $J_4<0$ is sufficient to induce in a discontinuous way the appearance of the intermediate phase, and a switch in passage probability density from $p(x_1,y)p(x_2,y)$ to (\ref{jointp}).

Also the passage to $J_4>0$ is discontinuous, since in this regime the vacua $++$ and $--$ become adjacent and the asymptotic profiles are given by (\ref{single17}) (Fig.~\ref{sigma1sigma2}). We then have an exact description of the wetting transition occurring at $J_4=0$ and associated to the interfacial tension 
\EQ
\Sigma_{++,--}=\left\{
\begin{array}{l}
2m\,,\hspace{1cm}J_4\leq 0\,,\\
2m\sin(\xi/2)\,,\hspace{1cm}J_4>0\,.
\end{array}
\right.
\EN
For $J_4>0$ there is formation of bubbles as in Fig.~\ref{bubble} with $a=++$, $b=--$, $c=+-$ or $-+$, $\alpha=\beta$ and $\gamma=\pi-\xi$. These bubbles contribute to the term (\ref{branching}) with coefficients which can be deduced from the results\footnote{In doing this one has to take into account that \cite{AT,DG} use the language of the high temperature phase, so that the operators $\sigma_i$ have to be replaced with their dual $\mu_i$.} of \cite{AT,DG}. Actually, only $\mathscr{C}_{++,--}^{\sigma_1\sigma_2}$ does not vanish, since $\langle\sigma_i(x,y)\rangle_{++,--}$ is odd in $x$ by symmetry and cannot include the even term (\ref{branching}); this is consistent with the fact that the contributions of the bubbles $+-$ and $-+$ to the profile of $\sigma_i$ have opposite sign and cancel each other.

\section{Conclusion}
In this paper we showed how field theory yields the exact asymptotic description of intermediate phases in the scaling limit of two-dimensional statistical systems at a first order phase transition point. The derivation is performed within the scattering formalism, in which the interfaces eventually correspond to the trajectories of kink excitations propagating in imaginary time. While this peculiarity of the two-dimensional case is intuitively clear, the technical way it enters the derivation is more subtle. In any dimensions the matrix elements of local operators contain disconnected parts corresponding to particles propagating without coupling to the operator. Only in two-dimensional space-time, however, the trajectory of a disconnected particle cannot be taken around the point where the local operator is inserted; it will pass either to the left or to the right of this point, a circumstance resulting into the presence of pole singularities in the connected part of the matrix element. These singularities determine the jumps in the order parameter which are the signature of phase separation.

A list of results we derived was given in the introduction and will not be repeated here. We only remark that the analysis was performed exploiting general properties of two-dimensional field theory at low energies. The information needed for specialization to models concerns the existence of bound states and the threshold values of kink-kink scattering amplitudes. The main models, however, are integrable in the scaling limit in two dimensions, and this information is available. In this way we were able, in particular, to establish the formation of an intermediate disordered phase in the dilute $q$-state Potts model, and to show how the Ashkin-Teller model yields an example of exactly solved bulk wetting transition. In all cases we determined the exact magnetization profiles and deduced from them the interface properties.

The analysis of this paper can be extended to cases in which, with the boundary conditions of Fig.~\ref{rectangle}, the interfacial region consists of more than two interfaces. This is expected, for example, in the regime III of the RSOS models \cite{ABF}, in which the degenerate vacua and the kinks connecting them form a chain in order parameter space. Scattering amplitudes \cite{LeClair} and matrix elements \cite{rsos} of the bulk theory are available, but a detailed study is beyond the scope of this paper.


\section*{Appendix}
This appendix is devoted to the derivation of the results
\bea
\label{digamma}
\digamma(x,y) &\equiv & \frac{1}{\zeta(R)} \int_{\mathbb{R}^{4}} \frac{\rd\theta_{1}\rd\theta_{2}\rd\theta_{3}\rd\theta_{4}}{(2\pi)^{4}} \, \frac{\theta_{12}^{2}\theta_{34}^{2}}{\theta_{13}\theta_{14}\theta_{23}\theta_{24}} \, \tilde{\mathcal{Y}}^{\star}(\theta_{1},\theta_{2},\theta_{3},\theta_{4})=\frac{\mathcal{G}(\chi) - 1Â}{2} \, , \\
\label{delta}
\Delta(x,y) &\equiv & \frac{i}{2\zeta(R)} \int_{\mathbb{R}^{4}} \frac{\rd\theta_{1}\rd\theta_{2}\rd\theta_{3}}{(2\pi)^{3}} \, \frac{\theta_{12}\theta_{13}}{\theta_{32}} \, \tilde{\mathcal{Y}}^{\star}(\theta_{1},\theta_{2},\theta_{3},\theta_{1})= \frac{\mathcal{L}(\chi)}{4} + \text{const} \, ,
\eea
where $\zeta(R)\equiv\frac{\text{e}^{-2mR}}{\pi(mR)^{2}}$, $\mathcal{G}$ and $\mathcal{L}$ were given in (\ref{double16}) and (\ref{L}), and
\EQ
\tilde{\mathcal{Y}}^{\star}(\theta_{1},\theta_{2},\theta_{3},\theta_{4})=\textrm{e}^{-2mR} \, \textrm{e}^{-\frac{mR}{4}\bigl[\theta_{1}^{2}+\theta_{2}^{2}+\theta_{3}^{2}+\theta_{4}^{2}\bigr] + imx \bigl[\theta_{3}+\theta_{4}-\theta_{1}-\theta_{2}\bigr]} \textrm{e}^{\frac{my}{2} \bigl[\theta_{3}^{2}+\theta_{4}^{2}-\theta_{1}^{2}-\theta_{2}^{2}\bigr]}
\label{expansion}
\EN
is (\ref{ystar}) evaluated at small rapidities. 
In the derivation we will use the Dawson function \cite{NIST}
\EQ
\label{dawsonf}
\mathcal{F}_{D}(x) \equiv \textrm{e}^{-x^{2}} \int_0^x \rd t \, \textrm{e}^{t^{2}} \,,
\EN
and the functions
\EQ
\omega_{n}(\lambda;a) = \int_{\mathbb{R}} \rd x \, \frac{x^{2n}}{x^{2}-a^{2}} \, \textrm{e}^{-\lambda x^{2}} \, ,
\label{omega_n}
\EN
with $n$ a non-negative integer. Let us evaluate (\ref{omega_n}). Using 
\EQ
\lim_{\epsilon\to0} \frac{1}{x-a \mp i\epsilon} = \pm \pi i \, \delta(x-a) + \mathcal{P}\frac{1}{x-a} \,,
\EN
we have
\EQ
\omega_{0}(\lambda;a) = \mathcal{P}\int_{\mathbb{R}} \frac{\rd x}{x^{2}-a^{2}} \, \textrm{e}^{-\lambda x^{2} } = \frac{1}{2a}\mathcal{P}\int_{\mathbb{R}} \frac{\rd x}{x-a} \, \textrm{e}^{-\lambda x^{2} } -\frac{1}{2a} \mathcal{P}\int_{\mathbb{R}} \frac{\rd x}{x+a} \, \textrm{e}^{-\lambda x^{2} } \, .
\EN
Defining
\EQ
\nu(\lambda,a) = \mathcal{P}\int_{\mathbb{R}} \frac{\rd x}{x-a} \, \textrm{e}^{-\lambda x^{2} }\,,
\EN
and using $x=u+a$ we have
\EQ
\partial_{a} \Bigl[ \nu(\lambda,a) \text{e}^{\lambda a^{2}} \Bigr] = -2\lambda \int_{\mathbb{R}} \rd u \, \textrm{e}^{-\lambda (u^{2}+2au) } = -2 \sqrt{\pi\lambda} \, \text{e}^{\lambda a^{2}} \,;
\EN
integrating with respect to $a$ and using $\nu(\lambda,0)=0$, we have
\EQ
\label{nu}
\nu(\lambda,a) = -2\sqrt{\pi} \, \text{erfi}(\sqrt{\lambda}a) \, \text{e}^{-\lambda a^{2}} = -2\sqrt{\pi} \mathcal{F}_{D}(\sqrt{\lambda} a) \, ,
\EN
where $\text{erfi}(z) = -i \text{erf}(iz)$ and $\mathcal{F}_{D}(z) = (\sqrt{\pi}/2) \text{erfi}(z) \, \text{e}^{-z^{2}}$ \cite{NIST}. It follows from (\ref{nu}) that
\EQ
\omega_{0}(\lambda;a) =  -2 \sqrt{\pi} \frac{\mathcal{F}_{D}(\sqrt{\lambda} a)}{a} \, ,
\EN
while for arbitrary $n$ we can use $\omega_{n}(\lambda;a) = \left( -\partial_{\lambda} \right)^{n} \omega_{0}(\lambda;a)$; in particular
\EQ
\omega_{1}(\lambda;a) = \mathcal{P}\int_{\mathbb{R}} \rd x \frac{x^{2}}{x^{2}-a^{2}} \, \textrm{e}^{-\lambda x^{2} } = \sqrt{\pi} \biggl[ \frac{1}{\sqrt{\lambda}}-2a\mathcal{F}_{D}(\sqrt{\lambda} a) \biggr] \, .
\EN

We will also need the result
\EQ
\label{dig05}
\Xi(\ell) \equiv \int_{\mathbb{R}} \rd u \, \frac{\mathcal{F}_{D}(u)}{u} \, \textrm{e}^{-u^{2}-i \ell u } = \frac{\pi^{3/2}}{4} \Biggl[ 1- \textrm{erf}^{2}\left( \frac{\ell}{\sqrt{8}} \right)\Biggr] \, ,
\EN
which can be derived as follows. Using the integral representation of the Dawson function
\EQ
\label{daw_rep}
\mathcal{F}_{D}(x) = \int_{0}^{\infty} \rd u \, \textrm{e}^{-u^{2}} \, \sin(2ux) \, ,
\EN
 we can write $\Xi(\ell)$ in the form
\EQ
\Xi(\ell) = \int_{0}^{\infty} \rd u \, Q(\ell,u) \, \textrm{e}^{-u^{2}} \, ,
\hspace{1cm}
Q(\ell,u) = \int_{\mathbb{R}} \rd x \, \frac{\sin(2ux)}{x} \, \textrm{e}^{-x^{2}-i\ell x} \, .
\EN
Taking the first derivative with respect to $\ell$ and carrying out the Gaussian integrations we find\footnote{Using the identity $\int_{\mathbb{R}} \rd x \frac{\sin(2ux)}{x} \, \textrm{e}^{-x^{2}} = \pi \, \textrm{erf}(u)$, we found that the integration constant is zero.} 
\EQ
\Xi(\ell) = \frac{\pi}{2} \int_{0}^{\infty} \rd u \, \textrm{e}^{-u^{2}} \, \Biggl[ \textrm{erf}\left( u - \frac{\ell}{2}\right) + \textrm{erf}\left( u + \frac{\ell}{2}\right) \Biggr] \, ,
\EN
and with the aid of
\EQ
\int_{0}^{\infty} \rd u \, \textrm{e}^{-u^{2}} \, \textrm{erf}(u+a) = \frac{\sqrt{\pi}}{4} \Bigl[ 2 - \textrm{erfc}^{2}\left( a/\sqrt{2} \right)\Bigr] \, ,
\EN
we obtain (\ref{dig05}).

Consider now (\ref{digamma}). Introducing the variables $x_{\pm} = \theta_{1} \pm \theta_{3}$, $y_{\pm} = \theta_{2} \pm \theta_{4}$, and then $u_{\mp} = x_{+} \mp y_{+}$, $v_{\mp} = x_{-} \mp y_{-}$, we have $\bigg\vert \det \frac{\partial \left( \theta_{1}, \theta_{2}, \theta_{3}, \theta_{4} \right)}{\partial \left( u_{-}, u_{+}, v_{-}, v_{+} \right)} \bigg\vert = \frac{1}{16}$, $\rd\theta_{1}\rd\theta_{2}\rd\theta_{3}\rd\theta_{4} = 16^{-1} \rd u_{+} \rd u_{-} \rd v_{+} \rd v_{-}$, and
\EQ
\tilde{\mathcal{Y}}^{\star}(\theta_{1},\theta_{2},\theta_{3},\theta_{4}) =\textrm{e}^{-2mR} \, \textrm{e}^{-\frac{mR}{16}\bigl[u_{+}^{2}+u_{-}^{2}+v_{+}^{2}+v_{-}^{2}\bigr] - imxv_{+}} \, \textrm{e}^{-\frac{my}{4} \bigl[u_{+}v_{+}+u_{-}v_{-}\bigr]} \,.\nonumber
\EN
Rescaling the variables as $u_\pm\rightarrow (4/\sqrt{mR}) u_\pm$ and $v_\pm\rightarrow (4/\sqrt{mR}) v_\pm$, we have $\digamma(x,y)=\pi^{-3} E(h,\epsilon)$, where
\EQ
E(h,\epsilon) \equiv \int_{\mathbb{R}^{4}} \rd u_{+} \rd u_{-} \rd v_{+} \rd v_{-} \, \frac{\left( u_{-}^{2}-v_{-}^{2} \right)^{2}}{\left( v_{+}^{2}-u_{-}^{2} \right) \left( v_{+}^{2}-v_{-}^{2} \right)} \, \textrm{e}^{ -\left( u_{+}^{2} + u_{-}^{2} +v_{+}^{2} +v_{-}^{2} \right) - i h v_{+} -2\epsilon (u_{+}v_{+}+u_{-}v_{-})} \, ,
\EN
with $h=4mx/\sqrt{mR}$ and $\epsilon=2y/R$. We write
\bea \nonumber
E(h,\epsilon) & = & \int_{\mathbb{R}^{2}} \rd u_{+} \rd v_{+} \, \textrm{e}^{-u_{+}^{2}} \, \textrm{e}^{-v_{+}^{2}-ihv_{+}-2\epsilon u_{+}v_{+}} \, f(v_{+},\epsilon) \\
& = & \sqrt{\pi}\int_{\mathbb{R}} \rd v_{+} \, \textrm{e}^{-\kappa^{2} v_{+}^{2}-ihv_{+}} \, f(v_{+},\epsilon) \, ,
\eea
where we have used the parameter $\kappa^{2}=1-\epsilon^{2}$ and the function
\bea
f(\alpha,\epsilon) &\equiv & \int_{\mathbb{R}^{2}} \rd x \rd y \, \frac{(x^{2}-y^{2})^{2}}{(x^{2}-\alpha^{2})(y^{2}-\alpha^{2})} \, \textrm{e}^{-x^{2}-y^{2}-2\epsilon x y} \nonumber\\
& = & \int_{\mathbb{R}^{2}} \rd x \rd y \, \frac{\Bigl[(x^{2}-\alpha^{2})-(y^{2}-\alpha^{2})\Bigr]^{2}}{(x^{2}-\alpha^{2})(y^{2}-\alpha^{2})} \, \textrm{e}^{-x^{2}-y^{2}-2\epsilon x y}\nonumber \\
& = & 2\int_{\mathbb{R}^{2}} \rd x \rd y \, \Biggl[ \frac{x^{2}-\alpha^{2}}{y^{2}-\alpha^{2}} - 1 \Biggr] \, \textrm{e}^{-x^{2}-y^{2}-2\epsilon x y}
= \sqrt{\pi} \int_{\mathbb{R}} \rd y \, \frac{1-2\kappa^{2}y^{2}}{y^{2}-\alpha^{2}} \, \textrm{e}^{-\kappa^{2}y^{2}} \,;
\eea
the integration over $y$ can be performed using the functions (\ref{omega_n}), thus
\EQ
f(\alpha,\epsilon) = \sqrt{\pi} \biggl[ \omega_{0}(\kappa^{2},\alpha) -2 \kappa^{2} \omega_{1}(\kappa^{2},\alpha) \biggr]  \, ,
\EN
and
\EQ
\label{dig03}
E(h,\epsilon) = 2\pi^{3/2}\kappa \, \int_{\mathbb{R}} \rd  u \, \textrm{e}^{-\kappa^{2}u^{2}-ihu} \biggl[ 2\kappa u \mathcal{F}_{D}(\kappa u) -\frac{\mathcal{F}_{D}(\kappa u)}{\kappa u} - 1 \biggr] \,.
\EN
The function (\ref{dig03}) satisfies $E(h,\epsilon) = E(h/\sqrt{1-\epsilon^{2}},0)$, as one can easily verify with the rescaling $\kappa u \rightarrow u$. We recall that $h/\kappa=\sqrt{8}\chi$, therefore (\ref{dig03}) becomes
\EQ
\label{dig04}
E(h/\kappa,0) = 2\pi^{3/2} \, \int_{\mathbb{R}} \rd  u \, \textrm{e}^{-u^{2}-i \sqrt{8}\chi u} \biggl[ 2 u \mathcal{F}_{D}(u) -\frac{\mathcal{F}_{D}(u)}{u} - 1 \biggr] \, .
\EN
It follows from (\ref{dawsonf}) that $\mathcal{F}_{D}^{\prime}(x) = 1 - 2x \, \mathcal{F}_{D}(x)$; using this property, an integration by parts of (\ref{dig04}) and recalling (\ref{dig05}) we can show that (\ref{dig04}) can be written in the form
\EQ
\label{dig06}
E(h/\kappa,0) = 2\pi^{3/2} \biggl[ \frac{1}{4}\partial_{\chi}^{2} + \chi\partial_{\chi} - 1 \biggr] \, \Xi(\sqrt{8}\chi)= \frac{\pi^{3}}{2}\biggl[ \mathcal{G}(\chi) - 1 \biggr] \, ,
\EN
which amounts to (\ref{digamma}).

We now turn to (\ref{delta}). We rescale the integration variables as $\theta_{i} \rightarrow \sqrt{2/mR} \, \theta_{i}$, perform the Gaussian integral over $\theta_{1}$ and for the remaining integration rapidities adopt the change of variables $\theta_{\pm} = \theta_{3}\pm\theta_{2}$; obtaining
\EQ
\Delta(x,y)=\frac{i}{32\pi^{3/2}} \int_{\mathbb{R}^{2}} \rd\theta_{+}\rd\theta_{-} \, \Bigl[ \frac{2 + \theta_{+}^{2} - \theta_{-}^{2}}{\theta_{-}} \Bigr] \, \text{e}^{-\frac{\theta_{+}^{2}+\theta_{-}^{2}}{4}+\frac{\epsilon}{2}\theta_{+}\theta_{-} + i (x/\lambda) \theta_{-}} \,;
\EN
Gaussian integration gives
\EQ
\label{del1}
\Delta(x,y)=\frac{i}{16\pi}\int_{\mathbb{R}} \frac{\rd \theta_{-}}{\theta_{-}} \, \left( 4- \theta_{-}^{2} \right) \, \textrm{e}^{-\frac{\theta_{-}^{2}}{4}+i \chi \theta_{-}} \,,
\EN
and then
\EQ
\partial_{\chi}\Delta(x,y)=-\frac{1}{4\sqrt{\pi}} (1+2\chi^{2}) \, \textrm{e}^{-\chi^{2}}.
\EN
Integrating back we obtain (\ref{delta}).


\begin{thebibliography}{99}
\bibitem{Dietrich} S. Dietrich, in Phase Transitions and Critical Phenomena, edited by C. Domb and J.L. Lebowitz,  Vol. 12, p.~1, Academic Press, London, 1988.

\bibitem{Abraham} D.B. Abraham, in Phase Transitions and Critical Phenomena, edited by C. Domb and J.L. Lebowitz,  Vol. 10, p.~1, Academic Press, London, 1986.

\bibitem{Fisher} M.E.~Fisher, J. Stat. Phys. 34 (1984) 667.

\bibitem{DV} G.~Delfino and J.~Viti, {J. Stat. Mech.} (2012) P10009.

\bibitem{DS} G.~Delfino and A. Squarcini, J. Stat. Mech. (2013) P05010.

\bibitem{BKW} B. Berg, M. Karowski and P. Weisz, Phys. Rev. D 19 (1979) 2477.

\bibitem{Smirnov} F.A. Smirnov, Form Factors in Completely Integrable Models of Quantum Field Theory, World Scientific, 1992.

\bibitem{DC98} G.~Delfino and J.~Cardy, {Nucl.Phys.} B 519 (1998) 551.

\bibitem{DV10} G. Delfino and J. Viti, Nucl. Phys. B 840 (2010) 513.

\bibitem{Abraham81} D.B. Abraham, {Phys. Rev. Lett.} \textbf{47} (1981) 545.

\bibitem{GI} L. Greenberg and D. Ioffe, Ann. Inst. H. Poincar\'e Probab. Statist. 41 (2005) 871.

\bibitem{CIV} M. Campanino, D. Ioffe and Y. Velenik, Ann. Probab. 36 (2008) 1287.

\bibitem{Wu} F.Y. Wu, Rev. Mod. Phys. 54 (1982) 235.

\bibitem{Baxter} R.J. Baxter, Exactly Solved Models in Statistical Mechanics, Academic Press, New York, 1982.

\bibitem{paraf} G. Delfino, Annals of Physics, 333 (2013) 1. 

\bibitem{CZ} L. Chim and A.B. Zamolodchikov, Int. J. Mod. Phys. A 7 (1992) 5317.

\bibitem{dilute} G.~Delfino, {Nucl. Phys.} B 554 (1999) 537.

\bibitem{FK}  C.M. Fortuin and P.W. Kasteleyn, J. Phys. Soc. Jpn. Suppl. 26 (1969) 11; Physica 57 (1972) 536.

\bibitem{DCq4} G. Delfino and J. Cardy, Phys. Lett. B 483 (2000) 303.

\bibitem{McCoyWu} B. M. McCoy and T.T. Wu, Phys. Rev. D 18 (1978) 1243.

\bibitem{AU} D.B.~Abraham and P.J.~Upton, {Phys. Rev. Lett.} 70 (1993) 1567.

\bibitem{KF} L.-F. Ko and M.E.~Fisher, {J. Stat. Phys.} 58 (1990) 249.

\bibitem{DG} G. Delfino and P. Grinza, Nucl. Phys. B 682 (2004) 521.

\bibitem{Kadanoff} L.P. Kadanoff and A.C. Brown, Ann. Phys. 121 (1979) 318.

\bibitem{ZZ} A.B. Zamolodchikov and Al.B. Zamolodchikov, Ann. Phys. 120 (1979) 253.

\bibitem{AT} G. Delfino, Phys. Lett. B 450 (1999) 196.

\bibitem{ABF} G. Andrews, R. Baxter and J. Forrester, J. Stat. Phys. 35 (1984) 193.

\bibitem{LeClair} A. LeClair, Phys. Lett. B 230 (1989) 103.

\bibitem{rsos} G. Delfino, Nucl. Phys. B 583 [FS] (2000) 597.

\bibitem{NIST} N. M. Temme, in NIST Handbook of Mathematical Functions,
edited by F.W.J. Olver, D.W. Lozier, R.F. Boisvert, C.W. Clark, Cambridge University Press (2010).


\end{thebibliography}
\end{document}